\documentstyle[eqsecnum,preprint,aps]{revtex}
\tightenlines

\newcommand{\be}{\begin{equation}}

\newcommand{\ee}{\end{equation}}
\newcommand{\bs}{\begin{mathletters}} 
\newcommand{\es}{\end{mathletters}} 
\newcommand{\ba}{\bs\begin{eqnarray}}
\newcommand{\ea}{\end{eqnarray}\es}

\newcommand{\bt}[1]{\bs\label{#1}\begin{eqnarray}}
\newcommand{\et}{\end{eqnarray}\es}

\newcommand{\figlab}[2]{\begin{figure}\caption{#2}\label{#1}\end{figure}}

\newcommand{\der}[2]{\frac{\partial #1}{\partial #2}} 
\newcommand{\eq}[1]{Eq.~(\ref{#1})}
\newcommand{\eqs}[1]{Eqs.~(\ref{#1})}
\newcommand{\comm}[2]{\left[#1, #2\right]}

\newcommand{\acomm}[2]{\left[a_{#1}, a_{#2}\right]}
\newcommand{\atcomm}[3]{\left[a_{#1}, \left[a_{#2}, a_{#3}\right]\right]}

\newcommand{\Acomm}[2]{\left[A_{#1}, A_{#2}\right]}

\newcommand{\threematrix}[9]{\left(\begin{array}{ccc}
#1 & #2 & #3 \\
#4 & #5 & #6 \\
#7 & #8 & #9 
\end{array}\right)}

\newcommand{\paper}[6]{#1 , #2 #3 {\bf #4} , #5 (19#6)}
\newcommand{\book}[5]{#1 , {\it #2} (#3, #4, 19#5)}

\newcommand{\refsec}[1]{Sec.~\ref{#1}}
\newcommand{\refapp}[1]{Appendix~\ref{#1}}
\newcommand{\reffig}[1]{Fig.~\ref{#1}}

\begin{document}

\preprint{ULDF-TH-2/10/97}

\title{Algebraic methods in the study of systems 
of the reaction-diffusion type} 

\author{M. Beccaria, G. Soliani} 

\address{Dipartimento di Fisica dell'Universit\`a, 73100 Lecce, Italy, and
Istituto Nazionale di Fisica Nucleare, Sezione di Lecce, Italy}

\maketitle

\begin{abstract}
Nonlinear systems of the reaction-diffusion type, including Gierer-Meinhardt 
models of autocatalysis, are studied by using Lie algebras coming from the 
prolongation structure. The consequences of this analytical approach, as
the determination of special exact solutions, are compared with the
corresponding results obtained via numerical simulations.
\end{abstract}
\pacs{}

\section{Introduction}

Models for biological pattern formation may be described by
reaction-diffusion (RD) equations of the type~\cite{Koch}
\be
\label{classe}
u_{kt} = \alpha_k \nabla^2 u_k+R_k(u),
\ee
(no sum over $k$), where $u=\{u_k\}$ are the dynamical fields, $k=1,2,\dots,
N$, $\alpha_k$ are constants and $R_k(u)$ are functions of the fields which
characterize the reactions among them.

The class of \eq{classe} includes popular models of chemical
reactions~\cite{Gray}. In
particular, when the diffusion coefficients $\alpha_k$ are not all equal,
Turing instabilities may occur and complicated patterns emerge which have
been related to morphogenesis, reaction front dynamics, self-organization
and so forth~\cite{Misc}.

Models of pattern generation in complex organisms are investigated mostly
via computer simulations. A minor attention is payed to the analysis of the
mathematical properties of the underlying equations. The knowledge of these
properties, combined with numerical studies, could be of help both to
compare theory and experiment, to check the validity of the models, and to
suggest possible improvements.

Keeping in mind this programme, in this paper we apply the prolongation
technique~\cite{ref4} to the $1+1$ dimensional RD system
\be
\label{evolution}
u_{kt} = \alpha_k u_{kxx}+R_k(u),
\ee
($k=1,2,\dots, N$), where first the reaction functions $R_k(u)$ are not
specified. Our approach contemplates the use of the concept of
{\it pseudopotential},
that is an $M$-component vector ($M$ arbitrary)
$y=\{y(x,t)\} = (y_1,\dots, y_M)$ defined by
\bt{linearized}
y_x &=& f^j(u) T_j y , \\
y_t &=& g^j(u,u_x) T_j y , 
\et
where summation over repeated indices is understood, 
$u=\{u_k\}$, $u_x=\{u_{kx}\}$, $T_j$ are $M\times M$ matrices
spanning a Lie algebra $\cal L$
\be
\left[T_i, T_j\right] = c_{ij}^k T_k,
\ee
and $f^j$, $g^j$ are functions to be determined in such a way that the
compatibility condition $y_{tx}=y_{xt}$ reproduces \eq{evolution}. When
this feature occurs, then \eqs{linearized} represent a linearization of
\eq{evolution}. 

The motivation for dealing with \eq{evolution} instead of \eq{classe}
within the prolongation scheme is at least threefold. First, the prolongation
machinery in $2+1$ dimensions is not yet well established, even if some
examples exist concerning a few integrable cases~\cite{Morris}. Second,
biological structures modeled by equations of the class~(\ref{evolution})
are interesting also in $1+1$ dimensions. One of them, which consists of
\eqs{evolution} where $R_k(u)$ is assumed to be linear in the fields
$\{u_k\}$ $(k=1,2)$, has been discussed recently by Kondo and Asai in relation
to the stripe pattern mechanism of the angelfish {\it
Pomachantus}~\cite{Kondo}. Third, the boundary of a complex two-dimensional
structure hosts a restricted $1+1$ dimensional RD system~\cite{Fassler}.  

The main results achieved in this work are listed in correspondence of the
contents of the different Sections in which the paper is planned.

In \refsec{sec:prolongation}, we deal with
the prolongation of RD systems described by
\eq{evolution}.
In \refsec{sec:quadratic}, we study a class of quadratic models which 
are completely linearizable and provide several exact analytical 
solutions of the travelling wave type. These solutions are obtained by direct 
inspection.
In \refsec{sec:linearizable}, we emphasize the role of
linearizability. The evolution equations for the pseudopotential are
written explicitly and solved showing how new solutions can emerge in 
terms of known ones.
In \refsec{sec:gm}, we apply the general results found in
\refsec{sec:prolongation} to the prolongation of 
the Gierer-Meinhardt models~\cite{Koch}. 
The resulting algebra turns out to be closed
and is that of the similitude group in the plane~\cite{Bluman}. 
In \refsec{sec:solutions}, we explore the consequences of this underlying
algebraic structure.
In particular, we treat analytical approximations of particular
classes of solutions, namely homogeneous and of the travelling wave type.
Section~\ref{sec:numerical} contains 
results obtained via numerical integration of the
evolution equations in order to check the analytical predictions and to 
suggest possible developments of the analysis.
Finally, in \refsec{sec:conclusions}
some concluding remarks are considered, while 
in the Appendices details of the calculations are reported.

\section{Prolongation of general RD systems}
\label{sec:prolongation}

Following the general strategy outlined in the Introduction, we begin our
analysis of the constraints on the algebra $\cal L$ by writing 
explicitly the compatibility condition $y_{tx}=y_{xt}$ of \eqs{linearized},
namely
\be
\label{compatibility}
\sum_k \left\{
\der{f^\gamma}{u_k}u_{kt}-\der{g^\gamma}{u_k}u_{kx}-
\der{g^\gamma}{u_{kx}}u_{kxx}
\right\}+c_{\alpha\beta}^\gamma f^\alpha g^\beta = 0 .
\ee
Substitution from \eq{evolution} into \eq{compatibility} yields
\be
\label{compatibility2}
\der{f^\gamma}{u_k}(\alpha_k u_{kxx}+R_k)-\der{g^\gamma}{u_k}u_{kx}-
\der{g^\gamma}{u_{kx}}u_{kxx}+c_{\alpha\beta}^\gamma f^\alpha g^\beta = 0 .
\ee
The coefficient of $u_{kxx}$ must vanish
\be
\alpha_k\der{f^\gamma}{u_k} = \der{g^\gamma}{u_{kx}} ,
\ee
so that
\be
g^\gamma = \alpha_k\der{f^\gamma}{u_k}u_{kx} + \Delta^\gamma(u) ,
\ee
where $\Delta^\gamma$ is a function of integration.
Putting this result back into \eq{compatibility2}, we find
\be
\der{f^\gamma}{u_k}R_k-\alpha_k\frac{\partial^2 f^\gamma}{\partial
u_k\partial u_l} u_{kx} u_{lx}-\der{\Delta^\gamma}{u_k} u_{kx} +
c^\gamma_{\alpha\beta} f^\alpha\alpha_k\der{f^\beta}{u_k}
u_{kx}+c^\gamma_{\alpha\beta} f^\alpha\Delta^\beta=0 .
\ee
Consequently, the following three conditions
\bt{threeconds}
(\alpha_k+\alpha_l) \frac{\partial^2 f^\gamma}{\partial
u_k\partial u_l} &=& 0, \\
\der{f^\gamma}{u_k}R_k
+c^\gamma_{\alpha\beta} f^\alpha\Delta^\beta &=& 0, \\
\der{\Delta^\gamma}{u_k} -c^\gamma_{\alpha\beta}
f^\alpha\alpha_k\der{f^\beta}{u_k} &=& 0 , \label{tmp1}
\et
hold. Equations~(\ref{threeconds}) entail two 
different cases distinguished by the condition
\be
\label{degeneracy}
\alpha_k+\alpha_l \neq 0 \qquad \forall k,l ,
\ee
or 
\be
\label{degeneracy2}
\alpha_k+\alpha_l = 0 \qquad  \mbox{for some}\ k, l.
\ee
Let us call ``non-degenerate'' and ``degenerate'' the problems
corresponding 
to \eq{degeneracy} and \eq{degeneracy2}, respectively. 
In these cases, we see that $f$ is linear in
the fields $u$ and, therefore, the structure of $f$ and $g$ may be
expressed by
\ba
f^\alpha &=& H^\alpha_k u_k + K^\alpha , \\
g^\alpha &=& \alpha_k H^\alpha_k u_{kx} + \Delta^\alpha(u) ,
\ea
where $H^\alpha_k$ and $K^\alpha$ are constants. Inserting these quantities
into \eq{tmp1} we obtain
\be
\der{\Delta^\gamma}{u_k} = c^\gamma_{\alpha\beta}(H^\alpha_l u_l +
K^\alpha)\alpha_k H_k^\beta = \alpha_k c_{\alpha\beta}^\gamma H^\alpha_l
H^\beta_k u_l + c^\gamma_{\alpha\beta} K^\alpha H^\beta_k \alpha_k .
\ee
On the other hand, the compatibility condition
\be
\frac{\partial^2 \Delta^\gamma}{\partial u_k\partial u_l} = 
\frac{\partial^2 \Delta^\gamma}{\partial u_l\partial u_k} ,
\ee
gives
\be
c^\gamma_{\alpha\beta} H^\alpha_l H^\beta_k \alpha_k = 
c^\gamma_{\alpha\beta} H^\alpha_k H^\beta_l \alpha_l 
\Rightarrow
(\alpha_k+\alpha_l) c^\gamma_{\alpha\beta} H^\alpha_k H^\beta_l = 0 .
\ee
From this relation, we get
\be
c^\gamma_{\alpha\beta} H^\alpha_k H^\beta_l = 0 
\ee
in the non-degenerate case.
Now we can exploit this equation to determine the $\Delta^\gamma$ term, i.e.
\be
\der{\Delta^\gamma}{u_k} = c^\gamma_{\alpha\beta} K^\alpha H_k^\beta
\alpha_k 
\Rightarrow
\Delta^\gamma = c^\gamma_{\alpha\beta} K^\alpha H_k^\beta \alpha_k +
D^\gamma ,
\ee
($D^\gamma$ being constants of integration), 
and the final expressions for $f$ and $g$, which turn out to be
\ba
f^\alpha &=& H_k^\alpha u_k + K^\alpha , \\
g^\alpha &=& \alpha_k H^\alpha_k u_{kx} + c^\alpha_{\beta\gamma} K^\beta
H^\gamma_k \alpha_k u_k + D^\alpha .
\ea
At this stage the reaction terms come into play.
All the unknown constants must be chosen in order to satisfy the
final equation
\be
\der{f^\gamma}{u_k}R_k + c_{\alpha\beta}^\gamma f^\alpha \Delta^\beta=0 .
\ee
To recognize the algebraic structure of the problem, let us 
adopt a notation in which the Lie algebra indices are understood.
We define
\bt{somedefinitions}
f &=& f^\alpha T_\alpha , \qquad g = g^\alpha T_\alpha , \\
K &=& K^\alpha T_\alpha , \qquad D = D^\alpha T_\alpha , \\
\Phi &=& H_k u_k =  H_k^\alpha u_k T_\alpha , \label{phipsidef} \\
\Psi &=& \alpha_k H_k u_k =  \alpha_k H_k^\alpha u_k T_\alpha , 
\et
and write the linearized problem in the compact form
\bt{structure}
f &=& \Phi+K , \\
g &=& \Psi_x + \comm K \Psi +D , 
\et
where
\bt{incomplete}
\comm {H_k}{H_l} &=& 0 , \\
\sum_k H_k R_k &+& \comm{\Phi+K}{\comm K \Psi+D} = 0 . \label{incomplete1}
\et
Equations~(\ref{incomplete}) 
must be satisfied whenever the fields $u$ are chosen. 
Taking the independent monomials in the fields, we
obtain an incomplete Lie algebra, in the sense that not all the commutators
are known. 
By construction, we know that, if the incomplete algebra is satisfied, then 
the compatibility condition for the linearized problem is assured as long as 
the evolution equations hold.

On the other hand, the explicit form
of the compatibility condition $y_{xt}=y_{tx}$ is easily found to be
\be
\label{generalcompatibility}
\Phi_t-\Psi_{xx} + \comm{\Phi+K}{\comm K \Psi+D} = 0 ,
\ee
which can be written as
\be
\label{reconstructed}
\Phi_t = \Psi_{xx} + \sum_k H_k R_k 
\ee
by virtue of \eq{incomplete1}.
If the elements $\{H_k\}$ of $\cal L$ were independent, then we could project
\eq{reconstructed} onto its components and recover the full set of
evolution equations estabilishing the complete linearization of the
reaction-diffusion system under consideration, i.e.
\be
\label{fullequiv}
y_{tx}=y_{xt} \leftrightarrow \mbox{evolution equations (\ref{evolution})}.
\ee
However, as we said before, $\{H_k\}$ are subject to the incomplete Lie
algebra~(\ref{incomplete}) which is a severe constraint. 
Actually, \eqs{incomplete} say that
every nonlinear term higher than a cubic polynomial in \eq{evolution},
implies a linear
dependence among the elements of the set $\{H_k\}$.
In other words, we have
\ba
\mbox{evolution equations} & \rightarrow & y_{tx}=y_{xt}, \\
y_{tx}=y_{xt} & \rightarrow & \mbox{linear combinations of evolution
equations,} \\
&& \mbox{at most quadratic in the fields} .
\ea
Therefore, the full equivalence expressed by \eq{fullequiv} may be achieved 
only if (i) the reaction terms are at most quadratic or (ii) in the special
degenerate cases where $f$ and $g$ are no more constrained to be linear in
the fields.
In all the other cases, \eqs{linearized} represent only a semi-linearization
of the evolution equations, i.e. they are equivalent only to a reduced set of 
linear combinations of the original equations.
For simplicity, also in these cases we shall 
keep calling \eqs{linearized} the {\it linearized problem} for \eq{evolution}.

We are interested in applications of the reaction-diffusion systems in 
problems related to morphogenesis, chemical autocatalysis and 
biological modeling. In these contexts negative diffusion coefficients 
do not admit a simple interpretation and therefore we shall not consider in
detail the degenerate case.
To this regard, we limit ourselves to mention the system
\bt{gravity}
u_{1t}-u_{1xx}+2u_1^2 u_2-2a u_1 &=& 0 , \\
u_{2t}+u_{2xx}-2u_1 u_2^2+2a u_2 &=& 0 , 
\et
where $a$ is a fixed constant. Equations~(\ref{gravity}), which admit an 
infinite
dimensional prolongation Lie algebra endowed with a loop
structure~\cite{Eleonora}, are integrable, and emerge in the gauge
formulation of the $1+1$ dimensional gravity, where $u_1$ and $u_2$ have
the meaning of Zweibein fields~\cite{Martina}. These equations are similar
to the ``fictitious'' or ``mirror-image'' systems with {\it negative}
friction, which appear into the thermo-field approach to the damped
oscillator~\cite{Celeghini}. Anyway, the role (if any) of \eqs{gravity} in
the modeling of complex organisms remains to be elucidated.

\section{Linearizable quadratic models}
\label{sec:quadratic}

Here we shall build up RD systems which are quadratic and linearizable. We
remind the reader that some of these models find applications in the study
of isothermal autocatalytic chemical reactions~\cite{Focant}. In this
context it is important the existence of propagating fronts (or travelling
waves) describing the advance of the reaction. Thus, the investigation of
RD systems allowing exact solutions of this kind can be a guide for the
construction of models which may interprete realistic chemical situations.

Let us start from a closed prolongation
algebra and study the explicit form of the compatibility condition 
expressed by  
$y_{xt}=y_{xt}$.
In the non-degenerate case, this condition is \eq{generalcompatibility},
namely a set of quadratic evolution equations, one for each independent 
Lie algebra
generator appearing after the expansion of the commutators.
Since $\Phi$ and $\Psi$ are the field dependent terms, the
quadratic, linear and constant terms are respectively
\be
\label{terms}
\comm \Phi {\comm K \Psi} , \qquad
\comm \Phi D +\comm K {\comm K \Psi} , \qquad
\comm K D .
\ee
Now, let us choose a definite Lie algebra $\cal L$. First, we fix 
the set of commuting elements $H=\{H_k\}$ involved in the definition of 
$\Phi$ and $\Psi$ (see \eq{phipsidef}). Second, we choose the
elements $K$ and $D$. Finally, we evaluate the quantities~(\ref{terms}).
The contributions proportional to the generators $H$ give genuine evolution
equations. All the other possible contributions fix constraints on the
fields.

If we look for non trivial systems, free of constraints, then we must
solve the following algebraic problem:
find a Lie algebra $\cal L$ such that: (i) $\cal L$ has an
abelian subalgebra $\cal A$ with dimension greater than 2 and (ii) given a
basis $\{H_i\}$ of $\cal A$ there exists $K\in\cal L$ such that 
\be
\comm{H_i}{\comm{H_j}{K}}\in \cal A .
\ee

In theory, we do not know if this problem admits solutions. 
For instance, if we assume $\cal L$ to be a semisimple algebra and identify
$\cal A$ with its Cartan subalgebra, then in the Cartan-Weyl basis~\cite{Gilmore} we have
\ba
\comm{H_i}{H_j} &=& 0, \\
\comm{H_i}{E_\alpha} &=& \alpha_i E_\alpha, \\
\comm{E_\alpha}{E_\beta} &=& N_{\alpha\beta} E_{\alpha+\beta}, \\
\comm{E_\alpha}{E_{-\alpha}} &=& \alpha^i H_i ,
\ea
and for a general $K$
\be
K = \sum_i \beta_i H_i + \sum_\alpha \gamma_\alpha E_\alpha ,
\ee
we obtain
\be
\comm{H_i}{\comm{H_j}{K}} = \sum_\alpha \alpha_i\alpha_j\gamma_\alpha 
E_\alpha , 
\ee
which does not belong to $\cal A$.

However, a simple class of algebras with the desidered property is provided
by the Ansatz
\be
\comm K D = 0, \qquad \comm K {\cal A} \sim D, \qquad \comm K D\in \cal A .
\ee
Indeed, application of the Jacobi identity determines all the commutators
which turn out to be
\bt{ilaria}
\comm{H_i}{H_j} &=& 0 , \ i,j = 1,\dots N\\
\comm D K &=& \gamma_i H_i , \\
\comm K {H_i} &=& \mu_i D, \\
\comm D {H_i} &=& \mu_i\lambda_j H_j , 
\et
where $\gamma$, $\mu$ and $\lambda$ are $N$ components vectors which must
satisfy
\be
\label{ortho}
\mu^T\lambda = 0 .
\ee
Now, let $u$ stand for the column vector of the fields and 
\be
A = \mbox{diag}(\alpha_1,\cdots, \alpha_N) .
\ee
Expanding the terms in \eq{terms} we get 
\be
\label{compactevolution}
u_t = Au_{xx}+(\lambda\ \mu^Tu+\gamma)(\mu^TAu+1) .
\ee
In the case $N=2$ it is natural to perform a change of 
variables and introduce in place
of $u_1$ and $u_2$, the new fields
\be
\label{change}
X = \mu^Tu,\qquad Y = \mu^T A u .
\ee
Exploiting the relation $\mu^T\lambda=0$ it is straightforward to show that
the two evolution equations for $(X,Y)$ are
\bt{XYevolution}
X_t &=& Y_{xx} + \mu^T\gamma (Y+1) ,\\
Y_t &=& (\alpha_1+\alpha_2) Y_{xx}-\alpha_1\alpha_2 X_{xx} +(\mu^TA\lambda\
X+\mu^T A\gamma)(Y+1).
\et
It is convenient to make the shift
\bt{shift0}
X &\to& X + \frac{\mu^T A\gamma}{\mu^T A\lambda}, \\
Y &\to& Y+1 ,
\et
in \eqs{XYevolution}. This yields
\bt{canonical}
X_t &=&  Y_{xx} + \mu^T\gamma Y, \label{firstcanonical}\\
Y_t &=& (\alpha_1+\alpha_2) Y_{xx}-\alpha_1\alpha_2  
X_{xx} + \beta(\alpha_1-\alpha_2) XY,
\et
where  
\be
\mu^T A \lambda = \mu_1\lambda_1(\alpha_1-\alpha_2)\equiv \beta
(\alpha_1-\alpha_2) ,
\ee
$\beta = \mu_1\lambda_1$ and \eq{ortho} has been used.
Now, let us look for solutions of the travelling wave type where both 
$X$ and $Y$ depend on $\xi=x+vt$. 
Furthermore, by taking $\mu^T\gamma=0$ and, consequently, 
$\gamma = \zeta \lambda$ ($\zeta$ being a constant factor), 
the shift in \eq{shift0} becomes simply
\bt{shift}
X &\to& X +\zeta,\\
Y &\to& Y+1 .
\et
Moreover, we can integrate the first of \eqs{canonical} and obtain
\be
\label{Xexpression}
X=\frac{1}{v} Y' + c_0,
\ee
where $Y'=dY/d\xi$ and $c_0$ is a constant of integration. 
The second equation reads
\be
vY'=(\alpha_1+\alpha_2) Y''-\frac{\alpha_1\alpha_2}{v}
Y'''+\frac{\beta}{v}(\alpha_1-\alpha_2) YY'+\beta c_0 Y,
\ee
which entails
\be
\label{2ndOrder}
Y''-v\left(\frac{1}{\alpha_1}+\frac{1}{\alpha_2}\right)
Y'+\frac{v^2}{\alpha_1\alpha_2}
Y-\frac{\beta}{2\alpha_1\alpha_2}(\alpha_1-\alpha_2) Y^2+c_1 = 0
\ee
for $c_0=0$, with $c_1$ arbitrary constant.

When the two diffusion constants $(\alpha_1, \alpha_2)$ take the special
values $(\alpha, 0)$ or $(\alpha, -\alpha)$ one of the coefficients in
\eq{2ndOrder} vanishes. Below, we shall discuss separately these singular
cases and the general one.

Anyhow, independently from the value of $(\alpha_1, \alpha_2)$, we can write
the following expressions for the fields $u_1$ and $u_2$ in terms of 
$X$ and $Y$ (see \eq{change}) and for the relative evolution equations
\ba
u_1 &=& \frac{1}{\mu_1(\alpha_2-\alpha_1)}(\alpha_2 X-Y), \\
u_2 &=& \frac{\lambda_2}{\lambda_1} \frac{1}{\mu_1(\alpha_2-\alpha_1)}
(\alpha_1 X-Y) , 
\qquad \mu_2 = -\frac{\lambda_1}{\lambda_2}\mu_1 , \\
&& \nonumber \\
u_{1t} &=& \alpha_1 u_{1xx} + \lambda_1 R(u_1, u_2), \\
u_{2t} &=& \alpha_2 u_{2xx} + \lambda_2 R(u_1, u_2), \\
R(u_1, u_2) &=& \left[\mu_1\left(u_1-\frac{\lambda_1}{\lambda_2}
u_2\right)+\zeta\right]\left[\mu_1\left(\alpha_1 u_1-\frac{\lambda_1}
{\lambda_2}\alpha_2 u_2\right)+1\right] .
\ea

\subsection{Case I: $\alpha_1=\alpha$, $\alpha_2=0$}

Equation~(\ref{2ndOrder}) becomes
\be
Y'=\frac{v}{\alpha} Y-\frac{\beta}{2v} Y^2,
\ee
which may be integrated to give
\be
Y(\xi) =
\frac{v^2}{\alpha\beta}\left(1+\tanh\frac{v}{2\alpha}(\xi-\xi_0)\right)  .
\ee
Computing $X$ from \eq{Xexpression} (with $c_0=0$), we find 
\bt{CaseI}
T &=& \tanh\frac{v}{2\alpha}(\xi-\xi_0) , \\
X &=& \frac{v^2}{2\alpha^2\mu_1\lambda_1}(1-T^2)-\zeta, \\
Y &=& \frac{v^2}{\alpha\mu_1\lambda_1}(1+T)-1 ,
\et
with the help of \eq{shift}.

\subsection{Case II: $\alpha_1=\alpha$, $\alpha_2=-\alpha$}

We have 
\be
Y''-\frac{v^2}{\alpha^2} Y+\frac{\beta}{\alpha} Y^2+c_1=0,
\ee
which furnishes
\be
\label{weier}
{Y'}^2=k_1+k_2 Y + \frac{v^2}{\alpha^2} Y^2-\frac{2}{3}\frac{\beta}{\alpha}
Y^3 
\ee
where $k_1$ and $k_2$ are arbitrary constants. The change of variable
\be
Y = -\frac{6\alpha}{\beta}\varphi+\frac{v^2}{2\alpha\beta},
\ee
leads to 
\ba
{\varphi'}^2 &=& 4\varphi^3-g_2\varphi-g_3, \\
g_2 &=& \frac{\beta}{6\alpha} k_2 + \frac{v^4}{12\alpha^4}, \\
g_3 &=& -\frac{\beta^2}{36\alpha^2} k_1-\frac{\beta v^2}{72\alpha^3} 
k_2-\frac{v^6}{216\alpha^6} ,
\ea
and therefore 
\be
Y(\xi) = -\frac{6\alpha}{\beta}{\cal P}(\xi-\xi_0, g_2, g_3) +
\frac{v^2}{2\alpha\beta} 
\ee
where $\cal P$ is the Weierstrass function~\cite{Abramowitz} and $\xi_0$ an 
arbitrary (complex) constant.
In the particular case $k_1=k_2=0$, we have $\Delta=g_2^3-27g_3^2=0$ 
which involves elementary functions only. Using
\be
{\cal P}(z, 12c^2, -8c^3) = c+\frac{3c}{\sinh^2(z\sqrt{3c})},
\ee
we find the following expressions for the fields $X$,  $Y$
\bt{CaseII}
X &=& -\frac{3v^2}{2\alpha^2\mu_1\lambda_1}
T(1-T^2)-\zeta, \\
Y &=& \frac{3v^2}{2\alpha\mu_1\lambda_1}(1-T^2)-1, \\
T &=& \tanh\frac{v}{2\alpha}(\xi-\xi_0) .
\et

\subsection{Case III: $\alpha_1$ and $\alpha_2$ arbitrary}

In general, \eq{2ndOrder} can be written as 
\be
Y'' = aY' + bY + c Y^2 + c_1,
\ee
where the coefficients $a$, $b$ and $c$ are defined by
\ba
a &=& v\left(\frac 1 {\alpha_1}+\frac 1 {\alpha_2}\right) ,\\
b &=& -\frac{v^2}{\alpha_1\alpha_2} ,\\
c &=& \frac{\beta}{2\alpha_1\alpha_2}(\alpha_1-\alpha_2) .
\ea
If we introduce the new dependent variable
\be
\tilde Y = Y-Y_0,
\ee
where 
\be
Y_0 = -\frac{1}{2c}\left(\frac 6 {25} a^2+b\right),
\ee
and choose $c_1$ to be
\be
c_1 = -b Y_0 - c Y_0^2,
\ee
then $\tilde Y$ satisfies 
\be
\label{mittag}
\tilde Y'' = a\tilde Y' -\frac 6 {25} a^2 \tilde Y + c \tilde Y^2.
\ee
In \refapp{app:mittag}, we show that the above equation affords the 
exact solution 
\be
\label{mittagsolution}
\tilde Y(\xi) = \exp\left(\frac{2a}{5}\xi\right){\cal P}\left(
\frac{5}{a}\sqrt\frac{c}{6}\exp\left(\frac{a\xi}{5}\right)+k_1, 0, k_0
\right),
\ee
($k_0$ and $k_1$ are constants of integration) corresponding to the 
socalled equianharmonic case~\cite{Abramowitz}.
For $k_0=0$, we deduce
\bt{CaseIII}
X &=& \frac{3a^3}{250 cv}(1-T)(1+T)^2-\zeta,\\
Y &=& \frac{3a^2}{50 c}(1+T)^2-1-\frac{1}{2c}
\left(\frac 6 {25} a^2+b\right),\\
T &=& \tanh\frac{a\xi}{10} ,
\et
where ${\cal P}(z,0,0)=z^{-2}$ has been exploited.

\subsection{Case IV: A particular solution for $\mu^T\gamma\neq 0$}

The condition $\mu^T\gamma = 0$ has been crucial in the previous
calculations because it has allowed the integration of 
\eq{firstcanonical}. In such a way, some exact solutions turn out to be 
polynomials in $\tanh(k\xi)$ for a certain $k$.
In the case $\mu^T\gamma\neq 0$, solutions of the similar kind can be
obtained by inserting the Ansatz
\ba
X &=& \sum_{n=0}^{N+1} a_n T^n, \qquad Y = \sum_{n=0}^{N} b_n T^n, \\
T &=& \tanh(k\xi) ,
\ea
into \eqs{canonical} ($N$ arbitrary).
The analysis of the case $N=2$, provides the solution
\bt{CaseIV}
X &=& \frac{3}{\rho_2 v}\left[(3D\rho_1-v^2)\sqrt{5\rho_1-\frac{v^2}{D}}
T-D\left(5\rho_1-\frac{v^2}{D}\right)^{3/2} T^3\right]-
\frac{\mu_1\gamma_1-\mu_2\gamma_2}{2\mu_1\lambda_1}, \nonumber \\
Y &=& -\frac{3 D}{\rho_2}\left(5\rho_1-\frac{v^2}{D}\right) (1-T^2)-1, \\
T &=& \tanh\left(\frac{\xi}{2}\sqrt{5\rho_1-\frac{v^2}{D}}\right),
\et
where 
\be
\alpha_1+\alpha_2=0, \ D = \alpha_1\alpha_2,\ \rho_1 = \mu^T\gamma,\ \rho_2
= \mu^T A\lambda .
\ee

In \reffig{fig:4exact}, we plot the fields $u_1(x,0)$ and $u_2(x,0)$ in 
the four
cases and with a choice of the parameters (specified in the captions) such that
reaction fronts arise. 

\section{Pseudopotential formulation of a
quadratic model}
\label{sec:linearizable}

In this Section, we formulate and solve the pseudopotential equations for a
particular linearizable quadratic model. We show that a bootstrap structure
emerges and new solutions can be obtained in terms of the old ones 
found by direct inspection. In doing so, let us assume 
\be
\lambda = (1,1), \ \gamma=(0,0), \ \mu = (1,-1),
\ee
in \eq{compactevolution}, which becomes the pair of evolution equations
\bt{pippo}
u_{1t} &=& \alpha_1 u_{1xx} + R(u_1,u_2), \\
u_{2t} &=& \alpha_2 u_{2xx} + R(u_1,u_2), 
\et
with
\be
R(u_1,u_2) = (u_1-u_2)(\alpha_1 u_1-\alpha_2 u_2+1).
\ee
In this case, the algebra~(\ref{ilaria}) reads
\bt{ilariapart}
\comm{H_1}{H_2} &=& \comm D K = 0,\\
\comm{K}{H_1} &=& -\comm{K}{H_2} = D, \\ 
\comm{D}{H_1} &=& -\comm{D}{H_2} = H_1+H_2.
\et
Up to now we have always interpreted the abstract elements 
$H_1$, $H_2$, $K$ and $D$ as matrices belonging to a given $N$-dimensional 
linear representation of the algebra. 
In such a case, the evolution equations for the
pseudopotential (see \eqs{somedefinitions} and (\ref{structure})) take the
form
\bt{matrixpp}
y_{ix} &=& F_{ij}(u) y_j, \\
y_{ix} &=& G_{ij}(u, u_x) y_j, 
\et
where $F_{ij}$ and $G_{ij}$ are field dependent matrices.
Hereafter, for convenience, we shall write \eqs{matrixpp} in the operator
form
\bt{vectorevolution}
{\cal Y}_x &=& {\cal F}, \\
{\cal Y}_t &=& {\cal G},
\et
where
\be
{\cal Y} = y_i\partial_i,\ {\cal F} =  F_{ij}(u) y_j\partial_i, \ 
{\cal G} =  G_{ij}(u) y_j\partial_i,
\ee
with $\partial_i = \partial/\partial y_i$.
In this way, one has associated with each abstract element 
$H_1$, $H_2$, $K$ and $D$ a differential operator (vector field)
 whose components, in the basis $\{\partial_i\}$, are linear
functions of the pseudopotential variables.
It is easy to see that this limitation is not necessary. Actually, all the
equations go unchanged if arbitrary vector fields are considered.
Taking this more general attitude, 
from the relation 
\be
\comm{\cal F}{\cal G} = {\cal F}{\cal G}-{\cal G}{\cal F} = 
-{\comm F G}_{ij} y_j\partial_i,
\ee
we see that the vector fields $H_1$, $H_2$, $K$ and $D$
satisfy the algebra
\bt{ilariapartbis}
\comm{H_1}{H_2} &=& \comm D K = 0,\\
\comm{K}{H_1} &=& -\comm{K}{H_2} = -D, \\ 
\comm{D}{H_1} &=& -\comm{D}{H_2} = -(H_1+H_2),
\et
which differs from \eqs{ilariapart} by a change of sign in the right hand side.
A possible realization of \eqs{ilariapartbis}
with a two component pseudopotential $y=(y_1,y_2)$ is
\ba
H_1 &=& \frac 1 8 \der{}{y_1}+\frac 1 2 \der{}{y_2}, \\
H_2 &=& \frac 1 8 \der{}{y_1}-\frac 1 2 \der{}{y_2}, \\
K   &=& \frac 1 2 y_2^2 \der{}{y_1}, \\
D   &=& \frac 1 2 y_2 \der{}{y_1}.
\ea
Hence, the pseudopotential equations~(\ref{vectorevolution}) take the form
\ba
y_{1x} &=& \frac 1 8 (u_1+u_2)+\frac 1 2 y_2^2, \label{pp1} \\
y_{2x} &=& \frac 1 2 (u_1-u_2), \label{pp2} \\
y_{1t} &=& \frac 1 8 (\alpha_1 u_{1x}+\alpha_2 u_{2x})+\frac 1 2 y_2
(1+\alpha_1 u_1-\alpha_2 u_2), \label{pp3} \\
y_{1t} &=& \frac 1 2 (\alpha_1 u_{1x}-\alpha_2 u_{2x}). \label{pp4}
\ea
Equations~(\ref{pp1})-(\ref{pp2}) produce
\bt{ufromy}
u_1 &=& 4y_{1x}+y_{2x}-2 y_2^2 , \\
u_2 &=& 4y_{1x}-y_{2x}-2 y_2^2 , 
\et
which can be used to eliminate $u_1$ and $u_2$ from
Eqs.~(\ref{pp3})-(\ref{pp4}). At this stage, let us seek
travelling wave solutions of the quadratic model~(\ref{pippo}), 
i.e. solutions of the
type $u_i=u_i(\xi)$ where $\xi=x+vt$ and $i=1,2$. To this aim, let us start
from \eq{pp4}, which can be written as
\be
\label{integration}
\frac{d}{d\xi}\left(-v y_1 + \frac 1 2 \alpha_1 u_1 - \frac 1 2 \alpha_2
u_2\right) = 0.
\ee
The integration of \eq{integration} gives
\be
\label{y1p}
{y'}_1(\xi)=\frac{1}{4(\alpha_1-\alpha_2)}\left[
2c_0-2vy_2+2(\alpha_2-\alpha_1)y_2^2+(\alpha_1+\alpha_2) y_2'
\right]
\ee 
with the help of \eqs{ufromy}, 
where $c_0$ is a constant of integration.
Taking account of \eq{y1p}, from \eq{pp3} we get 
\be
\label{determining}
y_2''-v\left(\frac 1 {\alpha_1}+\frac 1 {\alpha_2}\right) y_2' + 
v\left(\frac 1 {\alpha_2}-\frac 1 {\alpha_1}\right) y_2^2 + y_2\left[
\frac {v^2}{\alpha_1\alpha_2}+\left(\frac 1 {\alpha_2}-\frac 1 {\alpha_1}
\right)(2c_0-1)\right]-\frac{c_0 v}{\alpha_1\alpha_2}=0,
\ee
where only the pseudopotential component $y_2$ is involved.
This equation may be solved by the procedure exploited in Case III.
Precisely, we could make a shift $y\to y+y_0$ and fix the constants $y_0$,
$c_0$ in such a way that \eq{determining} takes the form \eq{mittag}.
However, for simplicity, here we adopt a different approach: we take
$c_0=0$ and fix $v$ by matching \eq{mittag}. Thus, we obtain
\be
\label{speed}
v = 5\sqrt\frac{\alpha_1\alpha_2(\alpha_2-\alpha_1)}
{(2\alpha_2-3\alpha_1)(3\alpha_2-2\alpha_1)},
\ee
which is real when
\be
\alpha_1 < \frac 2 3 \alpha_2, \qquad\mbox{or}\qquad 
\alpha_2 < \alpha_1 < \frac 3 2 \alpha_2 .
\ee
Just to furnish an explicit numerical example, let us consider $\alpha_1=2$,
$\alpha_2=3/2$. Formula~(\ref{speed}) gives $v=5$ and \eq{determining}
becomes
\be
\label{detpart}
y_2''(\xi)-\frac{35}{6}y_2'(\xi)+\frac{49}{6} y_2(\xi)-\frac{5}{6}y_2^2(\xi)=0.
\ee
As we have seen in Case III, a two-parameter solution of \eq{detpart} is 
\be
y_2(\xi) = \exp\left(\frac{7}{3}\xi\right){\cal P}
\left(\frac{\sqrt{5}}{7}\exp\left(\frac 7 6 \xi\right)+k_1, 0, k_0\right),
\ee
with $k_0$ and $k_1$ arbitrary constants. In particular, taking $k_0=0$
(see the discussion at the end of Case III) we obtain 
\be
y_2(\xi) = \frac{49}{20}\left(1+
\tanh\left(\frac {7}{12}(\xi-\xi_0)\right)\right)^2,
\ee
and, from \eqs{ufromy} and \eqs{y1p},
\ba
u_1 &=& \frac{49}{20}(1+T)^2(13+7T),\\
u_2 &=& \frac{49}{15}(1+T)^2(8+7T),\\
T &=& \tanh\left[\frac{7}{12}(x-x_0+5t)\right],
\ea
with $x_0$ arbitrary constant.

\section{The Gierer-Meinhardt models}
\label{sec:gm}

In this Section we apply the prolongation analysis to the RD models of the
Gierer-Meinhardt (GM) type. In the biological context, 
of a special interest is the cubic GM system, which 
describes the interplay between an {\it activator} field $u(x,t)$ and 
its {\it substrate} counterpart $w(x,t)$. 
This model is defined by the two equations  
\bt{cubicmodel}
u_t &=& \alpha u_{xx}+R_1,\\
w_t &=& \beta w_{xx} + R_2 ,
\et
with
\bt{cubic_interaction}
R_1 &=& \epsilon(u^2 w-u), \\
R_2 &=& \lambda(1-u^2 w) .
\et
The two fields diffuse with diffusion coefficients which are generally
different. The concentration of the field $u$ decays according to
the $-\epsilon u$ term, but is enhanced by the substrate $w$ via the
production term $\epsilon u^2 w$.
On the other hand, the substrate is injected in the domain of the reaction
with a constant rate $\lambda$ while its depletion is controlled by the
same nonlinear reaction term.

The two opposite fixed points are  
\be
\label{fixed}
\begin{array}{l}
u=1, \qquad w = 1, \\
u=0,\qquad w=\lambda t + \omega,\qquad \omega_t = \beta \omega_{xx} .
\end{array}
\ee
The former describes a scenario in which the activator field has reached 
a balance between catalyzation by substrate and substrate depletion.  
In the latter case the activator field is absent and the substrate arises
due to the injection constant term with inhomogeneities damped in time
according to the heat equation.

Moreover, 
according to the general analysis, it could be interesting also to
investigate the model obtained by replacing in \eqs{cubic_interaction}
the cubic interaction $u^2w$ with the quadratic term $uw$. We shall refer to 
this system as the quadratic Gierer-Meinhardt model.

\subsection{The cubic GM model}

Let us start from the prolongation equations
\ba
y_x &=& F(u,w,y), \\
y_t &=& G(u,u_x,w,w_x,y),
\ea
whose compatibility condition provides
\ba
\alpha F_u &=& G_{u_x}, \label{tmp3}\\
\beta  F_w &=& G_{w_x}, \label{tmp4}\\
F_u R_1+F_w R_2+\comm{F}{G} &=& G_u u_x + G_w w_x. \label{tmp2}
\ea
From these equations one finds that $G_{u_xu_x}=G_{w_xw_x}= 0$ 
and, therefore,
\be
G = A(u,w,y) u_x+B(u,w,y) w_x+C(u,w,y) ,
\ee
where $A$, $B$ and $C$ are vector fields of integration.
Putting back this result into \eq{tmp2} and equating to zero the 
coefficients of the independent monomials 
$1$, $u_x$, $w_x$, $u_xw_x$, $u_x^2$, $w_x^2$, we obtain
\bt{prolong}
A_u &=& 0 ,\\
B_w &=& 0 ,\\
A_w + B_u &=& 0 ,\\
\comm{F}{A} &=& C_u ,\\
\comm{F}{B} &=& C_w ,\\
F_u R_1 + F_w R_2 + \comm{F}{C} &=& 0 ,
\et
from which
\ba
A &=& a_0(y) w+a_1(y) ,\\
B &=& -a_0(y) u + a_2(y) , 
\ea
$a_0$, $a_1$ and $a_2$ being vector fields of integration.
Moreover, Eqs.~(\ref{tmp3}-\ref{tmp4}) imply
\be
\left\{\begin{array}{l}\alpha F_u=G_{u_x} ,\\
\beta F_w = G_{w_x} ,\end{array}\right. \Rightarrow
\left\{\begin{array}{l}
F = \frac{1}{\alpha} (a_0 uw+a_1 u + h(w,y)) ,\\
\frac{\beta}{\alpha}(a_0 u+ h_w) = -a_0 u + a_2 .
\end{array}\right.
\ee
Now, two cases have to be distinguished: if i) $\alpha+\beta \neq 0$, 
then
\ba
a_0 &=& 0 ,\\
h &=& \frac{\alpha}{\beta} a_2 w + a_3 ,
\ea
while if ii) $\alpha+\beta=0$, $a_0$ may be different from zero and
\be
h = -a_2 w + a_3 .
\ee
First let us analyze the non-degenerate case i) which is the one relevant to
proper RD systems~\cite{Koch}.

\subsubsection{The non-degenerate case: $\alpha+\beta\neq 0$}

The equations to be satisfied are (see \eqs{prolong})
\ba
\comm F A &=& C_u ,\label{fa}\\
\comm F B &=& C_w ,\label{fb}\\
F_u R_1 + F_w R_2 +\comm F C &=& 0 , \label{tmp7}
\ea
where
\ba
A &=& a_1, \qquad B = a_2 , \\
F &=& \frac{1}{\alpha}\left(a_1 u + \frac{\alpha}{\beta} a_2 w+a_3\right) ,
\\ 
G &=& a_1 u_x + a_2 w_x + C.
\ea 
Requiring that $C_{uw} = C_{wu}$ (see Eqs.(\ref{fa}-\ref{fb})), we find
\be
\acomm 1 2 = 0 ,
\ee
and 
\be
C = -\frac{1}{\alpha}\acomm 1 3 u -\frac{1}{\alpha} \acomm 2 3 w + a_4 .
\ee
Expanding \eq{tmp7} and collecting the coefficients of the monomials
$1$, $u$, $w$, $u^2$, $w^2$, $uw$ , $u^2w$ 
we are led to the following incomplete algebra
\ba
a_2 &=& \frac{\beta\epsilon}{\alpha\lambda} a_1 ,\\
\atcomm 1 1 3 &=& 0 ,\\
\acomm 3 4 &=& -\epsilon a_1 ,\\
\acomm 1 4 &=& \frac{1}{\alpha} \atcomm 3 1 3 + \epsilon a_1 ,\\
\acomm 1 4 &=& \frac{\beta}{\alpha^2} \atcomm 3 1 3 .
\ea
This algebra corresponds to \eqs{incomplete}.
The linear dependence between $a_1$ and $a_2$ is 
due to the cubic terms appearing in the evolution 
equations~(\ref{cubicmodel}). 
The compatibility condition $y_{tx} = y_{xt}$ 
together with the information encoded into the incomplete algebra 
gives 
\be
a_1 \left\{\mbox{first evolution eq.}\right\} \sim 
a_2 \left\{\mbox{second evolution eq.}\right\}  .
\ee
On the other hand, since $a_1$ and $a_2$ are linearly dependent,
this equation is not 
equivalent to the pair of evolution equations~(\ref{cubicmodel}).
However, some non trivial structure remains because the cubic terms in $R_1$
and $R_2$ are the same monomial in the fields.
If this had not been the case, then $a_1$ and $a_2$ would have been zero
with a corresponding trivial algebraic structure.

In order to close the above incomplete algebra, let us set
\be
\acomm 1 3 = a_5 .
\ee
If we introduce the parameter
\be
\xi^2 = \frac{\beta-\alpha}{\epsilon \alpha^2} ,
\ee
and rescale the generators as follows
\ba
a_1 &=& -\frac{\beta\epsilon\alpha^2}{(\beta-\alpha)^2}\xi A_1 ,\\
a_3 &=& \frac{1}{\xi} A_3 ,\\
a_4 &=& \frac{\epsilon\beta}{\beta-\alpha} A_4 ,\\
a_5 &=& -\frac{\beta\epsilon\alpha^2}{(\beta-\alpha)^2} A_5 ,
\ea
we have the commutation relations
\bt{quasisplit}
\Acomm 1 3 &=& A_5,\qquad \Acomm 1 4 = A_1 ,\\
\Acomm 1 5  &=& 0, \qquad \Acomm 3 4 = A_1 ,\\
\Acomm 3 5 &=& A_1, \qquad \Acomm 5 4 = A_5 ,
\et
in terms of $\{A_i\}$ $i=1,\dots 5$. Equations~(\ref{quasisplit}) define
a closed Lie algebra as one can check by computing all the non
trivial cases via the Jacobi identity.
Moreover, if we put
\be
A_6 = A_4-A_5 ,
\ee
then \eqs{quasisplit} become
\ba
\comm{A_1}{A_5} &=& 0,   \qquad   \Acomm 3 6 = 0 ,   \\
\Acomm 1 3 &=& A_5,      \qquad   \Acomm 1 6 = A_1 , \\
\comm{A_3}{A_5} &=& A_1, \qquad   \comm{X}{A_6} = A_5 ,
\ea
where we recognize the Lie algebra $sim(2)$ 
of the similitude group in the plane~\cite{Bluman}. 
The operators $A_1$ and $A_5$ are the generators 
of two orthogonal translations,
$A_3$ is the generator of rotations in the plane and, finally, $A_6$ 
is the generator of isotropic scalings. 
This geometrical interpretation gives also the basic vector field 
realization of the algebra, i.e.
\ba
A_1 &=& \der{}{x}, \qquad A_5 = \der{}{y} ,\\
A_3 &=& x\der{}{y}-y\der{}{x} ,\\
A_6 &=& x\der{}{x}+y\der{}{y} ,
\ea
It is rather intriguing the fact that the above rescaling requires
$\alpha\neq \beta$ since this is also a necessary condition for Turing
instability to occur in these kind of systems.
The point $\alpha=\beta$ is singular, in the sense that the incomplete
algebra~(\ref{incomplete}) collapses to 
\be
a_1=0,\qquad\acomm 3 4 = 0 ,
\ee
which implies the trivial result
\ba
y_x &=& \frac{1}{\alpha} a_3 ,\\
y_t &=& a_4, \\
\acomm 3 4 &=& 0.
\ea

\subsubsection{The degenerate case: $\alpha+\beta = 0$}

From \eqs{prolong} we deduce
\ba
F &=& a_0 uw+a_1 u-a_2 w + a_3 ,\\
G &=& A u_x + B w_x + C ,\\
A &=& \alpha (a_0 w + a_1) ,\\
B &=& \alpha (-a_0 u + a_2) ,\\
C &=& \alpha\left\{
uw\acomm 1 2 -u\acomm 1 3 -w\acomm 2 3 + a_4
\right\} .
\ea
We observe that in \eq{tmp7} the coefficient of the monomial $u^3 w$ must 
vanish and this gives $a_0 = 0$. 
Now, by performing the scaling $\epsilon\to \alpha\epsilon$, 
$\lambda\to\alpha\lambda$, the resulting prolongation algebra is 
\ba
\atcomm 2 2 1 &=& 0 ,\\
\atcomm 2 2 3 &=& 0 ,\\
\atcomm 1 1 3 &=& 0 ,\\
\atcomm 3 1 2 &=& 0 ,\\
\acomm 1 4 &=& \atcomm 3 1 3 + \epsilon a_1 ,\\
\acomm 2 4 &=& \atcomm 3 3 2 ,\\
\acomm 3 4 &=& \lambda a_2 ,\\
\atcomm 1 1 2 &=& -\epsilon a_1 -\lambda a_2 . \label{tmp9}
\ea
Here, in theory, the generators $a_1$ and $a_2$ are not constrained to be
linearly dependent. However, complete
linearization requires that specific algebraic constraints have 
to be satisfied.
For a generic model, some mechanism must ruin the equivalence between the 
evolution equations and the pseudopotential formulation.
In this case, as we promptly show, a linear dependence arises among the
generators.

From \eq{tmp9} we find
\ba
\epsilon \acomm 1 2 &=& \comm{a_2}{\atcomm 1 1 2} = \\
&=&  \comm{a_1}{\atcomm 2 1 2}-
\comm{\acomm 1 2}{\acomm 2 1} = 0 ,
\ea
hence, still from \eq{tmp9}, 
\be
a_2 = -\frac{\epsilon}{\lambda} a_1 ,
\ee
and the incomplete algebra turns out to close on the algebra $sim(2)$.

\subsection{The quadratic GM model}

In a similar way, also the quadratic Gierer-Meinhardt model turns out to be non
linearizable. All the details of the 
analogous computations are contained in \refapp{app:quadratic}.

\section{Special solutions of the non-degenerate cubic GM model}
\label{sec:solutions}

In this Section we focus on the cubic GM model in the non-degenerate case
and discuss the consequences of the pseudopotential formulation, which can
be written as 
\bt{cubic_lin}
y_x &=& Fy = -\left(u+\frac{\epsilon}{\lambda}w\right)\mu\xi
A_1 y+\frac{1}{\alpha\xi}A_3 y,\\
y_t &=& Gy = 
-\left(u+\frac{\beta\epsilon}{\alpha\lambda}w\right)_x\alpha\mu\xi
A_1 y+\left(u+\frac{\beta\epsilon}{\alpha\lambda}w\right)\mu A_5 y+\nu A_4 y,
\et
where
\be
\mu = \frac{\beta\epsilon\alpha}{(\beta-\alpha)^2},\qquad 
\nu = \frac{\epsilon\beta}{\beta-\alpha},\qquad 
\xi^2 = \frac{\beta-\alpha}{\epsilon\alpha^2} .
\ee
An explicit representation of the Lie algebra $sim(2)$ is obtained from the
geometric transformations of the similitude group and is given by
\be
A_1 = \threematrix
0 0 1
0 0 0 
0 0 0 , \qquad
A_3 = \threematrix
0 1 1 
{-1} 0 0 
0 0 0 
\ee
\be
A_4 = \threematrix
{-1} 0 0 
0 {-1} 0 
0 0 0 , \qquad
A_5 = \threematrix
0 0 0 
0 0 1
0 0 0 .
\ee
Since the third row is null, the third component of the pseudopotential is
constant, $y_3=c$. The other equations turn out to be
\bt{tmp12}
y_1 &=& -\alpha\xi y_{2x} ,\\
y_{2xx}+\frac{1}{(\alpha\xi)^2} y_2 &=& \frac{\mu
c}{\alpha}\left(u+\frac{\epsilon}{\lambda}w\right)-\frac{c}{(\alpha\xi)^2}
,\\
y_{2t} &=& \mu c \left(u+\frac{\epsilon\beta}{\lambda \alpha}w\right)-
\nu y_2 .
\et
The compatibility condition $y_{2,txx} = y_{2,xxt}$ gives the equation
\be
\label{linear}
\partial_t\left(u+\frac{\epsilon}{\lambda}w\right) = \alpha\partial^2_{x}
 \left(u+\frac{\epsilon\beta}{\lambda \alpha}w\right)+\epsilon(1-u) .
\ee
If we write $u$ and $w$ 
in terms of the pseudopotential $y\equiv y_2$ we find
\ba
u &=& 1+\frac{\beta-\alpha}{\epsilon\beta}\frac{1}{c}\left(\beta D_x
 -D_t \right)y ,\\
w &=& -\frac{\lambda\alpha}{\epsilon\beta}\left[
 1+\frac{\beta-\alpha}{\epsilon\alpha}\frac{1}{c}\left(\alpha D_x 
 -D_t \right)y\right] ,
\ea
where
\ba
D_x &=& \partial^2_{x} + \frac{1}{(\alpha\xi)^2} ,\\
D_t &=& \partial_t + \nu .
\ea
In order to determine $y$, we must put these
expressions into one of the genuinely nonlinear evolution equations. 
When this is done, one obtains
\be
\label{master}
\Omega(\alpha)\Omega(\beta) y =
\frac{\epsilon^2\beta}{\beta-\alpha}\left\{1+\frac{\alpha\lambda}
{\beta\epsilon}\left(1+\frac{\beta-\alpha}{\epsilon\beta}\Omega(\beta)y\right)^2
\left(1+\frac{\beta-\alpha}{\epsilon\alpha}\Omega(\alpha)y\right)\right\} ,
\ee
where $\Omega(\eta) = \eta D_x - D_t$ and in particular
$\Omega(\beta) = \beta\partial^2_{x}-\partial_t$.

In \refapp{app:casimir} we discuss the role of the Casimir operator ${\cal C} =
A_1^2 + A_5^2$ of the Euclidean subalgebra of $sim(2)$. In particular, we 
show that if $(u,w)$ is a solution of \eqs{cubicmodel} and $y$ a solution
of \eqs{cubic_lin}, then the new pseudopotential
\be
y\to  e^{-2\nu t}{\cal C} y
\ee
is a new solution.

Equation~(\ref{master}) will be applied below to find 
particular solutions (homogeneous and of the travelling wave type) 
to the cubic GM model.

\subsection{Homogeneous solutions}
\label{sec:homo}

Let us look for a class of solutions to \eqs{cubicmodel} $(\alpha\neq\beta)$
assuming that
\be
\label{Ansatz}
u+\frac{\epsilon}{\lambda} w = \phi(t),
\ee
where $\phi(t)$ denotes a given function of the time only.
Then, from \eqs{tmp12} we get
\be
u+\frac{\epsilon\beta}{\alpha\lambda}w =
\alpha\xi^2(\dot\phi+\nu\phi)-\frac{\nu}{\mu} = \cdots =
1-\frac\beta\alpha+\frac\beta\alpha\phi+\frac{\dot\phi}{\epsilon}
\left(\frac\beta\alpha-1\right),
\ee
and therefore
\bt{uwfromphi}
u &=& 1-\frac 1 \epsilon\dot\phi ,\\
w &=& \frac\lambda\epsilon\left(-1+\phi+\frac 1 \epsilon\dot\phi\right) ,
\et
where $\dot\phi=d\phi/dt$.
The equations of motion~(\ref{cubicmodel}) may be written as
\ba
u_t+\frac\epsilon\lambda w_t &=& \epsilon(1-u) ,\\
w_t &=& \lambda(1-u^2 w) .
\ea
The former is automatically satisfied by the Ansatz~(\ref{Ansatz}), 
while the latter may be
written as
\be
\phi'' = (1-\phi')\left(1+\frac\lambda\epsilon(1-\phi') 
(1-\phi-\phi')\right) ,
\ee
where $\phi'=d\phi/d\tau$, with $\tau=\epsilon t$.
The (trivial) solutions corresponding to $u=1$ and $u=0$  
are respectively
\be
\phi=1+\frac\epsilon\lambda,\qquad \phi=\tau-\tau_0 .
\ee
The problem to be solved is
\be
\label{a_def}
{\phi}'' = (1-\phi')\left(1+a(1-\phi') 
(1-\phi-\phi')\right),\qquad a = \frac\lambda\epsilon , 
\ee
with the initial conditions
\ba
\phi'(0) &=& 1-u(0) ,\\
\phi(0) &=& u(0)+\frac\epsilon\lambda w(0) .
\ea
Equation~(\ref{a_def}) can be elaborated by using 
the hodographic transformation 
\be
\label{hodographic}
\phi'=\frac 1 {1+\theta(\phi)} .
\ee
In doing so, we obtain
\be
\theta'=-\theta+\theta^2(a\phi-2)+\theta^3(a\phi-1-a),
\ee
which gives for $a=1$
\be
\label{thetaevolution} 
\theta' = \theta(\xi\theta(\theta+1)-1),
\ee
where $\theta'=d\theta/d\xi$, $\xi = \phi-2$ and
\be
\label{tmp14}
\phi'(\tau) = \frac 1 {1+\theta(\phi-2)}.
\ee
It is interesting to consider 
the qualitative behavior of \eq{thetaevolution} with $\theta(0)=\alpha>0$. 
This evolution problem 
\ba
\theta'(\xi) &=& -\theta+\xi\theta^2+\xi\theta^3 ,\\
\theta(0) &=& \alpha>0 ,
\ea
has solutions which are decreasing until they eventually meet the
nullcline $\Gamma$ given by 
$\theta(\theta+1)\xi=1$. It seems reasonable to predict
that all the
solutions starting with $\alpha$ below some critical value $\alpha^*$ decay
exponentially at infinity whereas all the other solutions meet $\Gamma$ and
thereafter explode in a finite time. This scenario, described in
\reffig{fig:qualitative},
is confirmed by the exact solution of the evolution problem when only one of 
the nonlinear terms is present, namely
\ba
\theta' &=& -\theta+\xi\theta^2 \Rightarrow 
\theta = \left(1+\xi+(1/\theta_0-1)e^\xi\right)^{-1} ,\\
\theta' &=& -\theta+\xi\theta^3 \Rightarrow 
\theta = \left(1+2\xi+(1/\theta_0^2-1) e^{2\xi}\right)^{-1/2} .
\ea 
A heuristic evaluation of $\alpha^*$ is shown in \refapp{app:heuristic}. 
Here we
describe a notable approximate solution to the evoution 
problem~(\ref{thetaevolution}). 
To this aim let us consider the equation
\be
\label{exact}
\xi\xi'-\frac{1}{\theta(\theta+1)} \xi'-\frac{1}{\theta^2(\theta+1)}=0 ,
\ee
with $\xi' = d\xi/d\theta$. Now we remind the reader that the class of
integrable equations
\be
\label{integrable}
\xi\xi'+f(\theta)\xi'\pm f'(\theta)=0 ,
\ee
allows the general integral
\be
\label{generalintegral}
I=e^{\pm \xi}(\xi+f(\theta)\mp 1) .
\ee
Actually, \eq{integrable} can be associated which the evolution equation
\be
\label{tmp13}
\theta'=\mp\frac 1 {f'(\theta)}(\xi+f(\theta)) ,
\ee
or
\be
\frac{df}{d\xi}=\mp(\xi+f) ,
\ee
which shows how integrability of \eq{integrable} is just linearity in
disguise. Anyhow, if we take
\be
f(\theta) = -\frac 1 {\theta(\theta+1)} ,
\ee
and the minus sign in \eq{tmp13}, then we find
\be
\label{modified}
\theta'=\frac{\theta(\theta+1)}{2\theta+1}
\left[\xi\theta(\theta+1)-1\right] .
\ee
This is a modified equation but with a modifying factor
$\frac{\theta(\theta+1)}{2\theta+1}$ which is bounded between 
$1/2$ and $1$ when $\theta>0$ and which, therefore, may be 
expected to give minor changes in the solution.
Another way of emphasizing the extra terms is that of writing 
\be
\label{extra}
\frac{d}{d\xi}\left[2\theta-\log(1+\theta)\right] = 
\frac{d}{d\xi}\left[\theta+\frac{\theta^2}{2}+\cdots\right] = 
\theta[\xi\theta(\theta+1)-1] .
\ee
The integral of \eq{modified} is (see \eq{generalintegral}):
\be
I=e^{-\xi}\left[\xi-\frac{1}{\theta(\theta+1)}+1\right] .
\ee
In terms of $\theta_0=\theta(0)>0$ we have 
\be
I=1-\frac{1}{\theta_0(\theta_0+1)} ,
\ee
and
\be
\theta(\xi) = -\frac 1 2 +\frac 1 2\sqrt{1+\frac 4{1+\xi-I e^\xi}} .
\ee
Some curves are displayed in \reffig{fig:modified}.
The constant $I$ is in the range $(-\infty, 1)$, all the solutions with
$I<0$ are regular for $\xi>0$ and decay exponentially at $\xi\to
+\infty$. The particular value $I=0$ corresponds to the separatrix between
regular and singular solutions and starts at the golden ratio 
$\theta_0 = (\sqrt 5-1)/2$. We
remark that \eq{tmp14} predicts singular solutions to be
associated with periodic oscillating solutions 
approaching limiting closed curves in the
$(u,w)$ plane as $t\to+\infty$.

In the special case of the separatrix, all the remaining integrations 
may be performed. At $I=0$ we have
\be
\theta(\xi)=-\frac 1 2 +\frac 1 2 \sqrt{1+\frac 4 {1+\xi}}\rightarrow
\theta(\phi-2)=-\frac 1 2 +\frac 1 2 \sqrt\frac{\phi+3}{\phi-1} ,
\ee
and hence integrating \eq{tmp14} we find
\be
\label{tmp15}
\frac 1 2 \phi+\frac 1 2 (\phi-1)
\sqrt{\phi+3}{\phi-1}+\log\left[\phi+1+(\phi-1)
\sqrt\frac{\phi+3}{\phi-1}\right]=\tau-\tau_0 .
\ee
Moreover, from \eqs{uwfromphi} 
\be
u(\tau)=1-\phi'=\cdots=\frac{\sqrt{(\phi+3)/(\phi-1)}-1}
{\sqrt{(\phi+3)/(\phi-1)}+1},
\ee
that is
\be
\label{phifromu}
\phi=\frac 1 u + u-1.
\ee
Finally, substitution from \eq{phifromu} into \eq{tmp15} gives
\be
\label{schroder}
-\frac{1}{2} + \frac{1}{u}+ \log \frac 2 u =\tau-\tau_0.
\ee
We remark that \eq{schroder} can be exactly solved in terms of the Lambert
$W$ function~\cite{Lambert} which obeys the Schr\"oder 
equation~\cite{Hagedorn}
\be
W(x) + e^{W(x)} = x.
\ee
Indeed, we have
\be
\label{lambert1}
u = \left[W\left(\frac 1 2 e^{\frac 1 2 + \tau-\tau0}\right)\right]^{-1} .
\ee
The function $w$ can be determined by \eqs{uwfromphi} with the help of
\eq{phifromu}, i.e. 
\be
\label{lambert2}
w = \frac{1}{u}-1 = W\left(\frac 1 2 e^{\frac 1 2 +
\tau-\tau0}\right)
-1 .
\ee
In \reffig{fig:lambert} we display the behavior of this special solution.
To conclude this Section, we point out that the asymptotic behavior of
$W(x)$ for large $x$ is $W(x)\sim \log x$ and therefore in this limit 
\ba
u(\tau) & \sim & \left(\tau-\tau_0+\frac 1 2 -\log 2\right)^{-1},\\
w(\tau) & \sim & \tau-\tau_0-\frac 1 2 -\log 2 .
\ea
We remark that this solution has been obtained from the approximate
equation~(\ref{modified}); however, it solves asymptotically the exact 
equations~(\ref{cubicmodel}) because the extra terms in \eq{extra} vanish
in the limit $u\to 0$ (see Eqs.~(\ref{uwfromphi}) and (\ref{hodographic})).

\subsection{Travelling waves}

In order to look for solutions of \eq{master} of the travelling wave type,
we assume that the pseudopotential $y$ is a function of the reduced variable
$\xi = x+vt$, where $v$ is a constant.
In terms of $\xi$ we have 
\ba
\Omega(\alpha) &=& \beta D^2-v D, \\
\Omega(\beta)  &=& \alpha D^2 -v D -\epsilon, 
\ea
with $D = \frac{d}{d\xi}$.
This formalism suggests the particular speed
\be
\label{critical}
v = \pm\beta\sqrt\frac{\epsilon}{\alpha-\beta},
\ee
as a critical value. Actually, in this case we have the factorization
\ba
\Omega(\alpha) &=& 
\alpha\left(D\pm\frac{\sqrt{\epsilon(\alpha-\beta)}}{\alpha}\right)
\left(D\mp\sqrt\frac{\epsilon}{\alpha-\beta}\right) ,\\
\Omega(\beta) &=& \beta D \left(D\mp\sqrt\frac{\epsilon}{\alpha-\beta}\right),
\ea
with a common factor. Therefore, we can set
\be
\varphi = \left(D\mp\sqrt\frac{\epsilon}{\alpha-\beta}\right) y ,
\ee
and get a reduced equation in terms of the variable $\varphi$. 
Without loss of generality, let us focus on the particular set of
constants
\be
\alpha=2,\quad \beta=1,\quad \epsilon = 1 .
\ee
Then, $v=\pm 1$ and the relation between $\varphi$ and $y$ becomes
$\varphi = (D\mp 1) y$. Equation~(\ref{master}) may be written as
\be
\label{mastercritical}
\left(D\pm 1/2\right)\left(D\mp 1\right) D\varphi = -\frac{1}{2}\left[
1-2\lambda\left(D\varphi-1\right)^2\left(\left(D\pm 1/2\right)
\varphi-1\right)\right] ,
\ee
and the fields $u$, $w$ are expressed in terms of $\varphi$ by 
\ba
u &=& 1-D\varphi ,\\
w &=& -2\lambda\left(1-\left(D\pm 1/2\right)\varphi\right).
\ea
The fixed points of the cubic Gierer-Meinhardt model (see \eq{fixed})
correspond to the two exact (trivial) solutions
\be
\label{fixedphi}
\varphi = \frac{1+2\lambda}{\lambda}, \qquad \varphi = \xi + \mbox{const}.
\ee
It is very interesting to look for a particular phenomenological meaning of
the critical value expressed by \eq{critical}. 
Indeed, our numerical simulations show that something special happens 
at that point (see \refsec{sec:numerical}).

Apart from the critical speed, the general problem of finding a travelling
wave solution to \eq{master} may be studied by looking for interpolating
solutions which at $\xi\to\pm\infty$ approach the two fixed
points~(\ref{fixedphi}). 
This is a difficult eigenvalue problem for a 4th order
nonlinear ordinary differential equation. 
We shall exhibit semi-analytical approximations of these 
interpolating solutions in \refsec{sec:numerical}, devoted mainly 
to numerical results. 

\section{Numerical Simulations}
\label{sec:numerical}

In this Section we discretize the evolution equations of the cubic GM model
and study them from a
numerical point of view. This allows for a check
of our analytical predictions and makes a step forward from our
approximate solutions toward the unknown exact ones and their
phenomenology.

\subsection{Homogeneous solutions}

Let us consider \eqs{cubicmodel} and (\ref{cubic_interaction}) in the case
$\lambda=\epsilon$ ($a=1$). By rescaling the time variable we can set 
\be
\lambda=\epsilon=1 .
\ee
The couple of differential equations to be integrated numerically is 
\bt{num1}
\dot u(t) &=& u^2 w-u ,\\
\dot w(t) &=& 1-u^2 w .
\et
The initial values $u(0)$, $w(0)$ are related by  
\be
w(0) = 2-u(0).
\ee
Since the theoretical analysis carried out in \refsec{sec:homo} shows that 
oscillating trajectories emerge for
\be
u(0) > \frac{\alpha^*}{1+\alpha^*} ,
\ee
where $\alpha^*$ is a critical value determined in \refapp{app:heuristic},
we shall start with $u(0)$ slightly above $\alpha^*/(1+\alpha^*)$. Using 
the following second order algorithm
for the integration of equations $\dot x_i = f_i(x)$:
\ba
y^{(n)}_i &=& x_i^{(n)} + \frac\delta 2 f_i(x^{(n)}) ,\\
x^{(n+1)}_i &=& x_i^{(n)} + \delta f_i(y^{(n)}) ,
\ea
with $\delta$ the discretization time step, we obtain the result pictured
in \reffig{fig:homo}.

\subsection{Travelling waves}

The search for travelling wave solutions of the cubic GM model amounts to 
the solution of a 4th order nonlinear ordinary differential equation
(a reduction of \eq{master}) with given boundary conditions corresponding
to the two fixed points~(\ref{fixed}). 
From a biological point of view, these solutions interpolate between 
regimes dominated by the activator field or the substrate field and are
characterized by a reaction front which separated the two regions.
For a 2nd order equation, phase-space techniques permit geometrical 
proofs of the existence 
of such solutions. Here, dealing with a 4th order equation
we shall build up approximate solutions and adopt them as  particular 
initial conditions in the numerical integration.

At the critical speed, the order is only 3 and we start from this case.
The equation to be solved is \eq{mastercritical}.
Without loss of generality, let us set $\lambda=1$. The two fixed 
points~(\ref{fixedphi}) are
\be
\varphi = \xi + \mbox{constant}, \qquad \varphi=3 .
\ee
As we shall see, the above constant can be set to zero. 
Let us look for a perturbed solution around $\varphi=\xi$ by writing
\be
\varphi = \xi+\delta ,
\ee
and expanding \eq{mastercritical} at the first order in $\delta$. 
We find
\be
2\delta^{'''}-\delta^{''}-\delta'=0 ,
\ee
whose characteristic roots are $0, 1, -1/2$.
The same computation starting from 
\be
\varphi = 3+\delta ,
\ee
gives 
\be
2\delta^{'''}-\delta^{''}-\delta'-\delta=0 ,
\ee
which has one characteristic root on the right complex half-plane and
the other two on the left half-plane.
A sensible perturbation must vanish at infinity. In order to have the 
greatest number of free constants we match the first fixed point at 
$\xi\to -\infty$ and the other at $\xi\to + \infty$.
Hence, we construct our approximate solution as 
\be
\varphi(\xi) = \left\{\begin{array}{ll}
\varphi_-(\xi) & \xi < 0 , \\
\varphi_+(\xi) & \xi \ge 0 , 
\end{array}\right.
\ee 
with 
\ba
\varphi_-(\xi) &=& \xi+c_1+c_2 e^\xi , \\
\varphi_+(\xi) &=& (c_3 \sin\nu\xi+c_4\cos\nu\xi) e^{-\mu\xi}+3 ,\\
\mu &=& 0.366876 ,\\
\nu &=& 0.520259 .
\ea
The constant $c_1$ may be eliminated and set to zero by a translation in
$\xi$. The free constants $c_2$, $c_3$ and $c_4$ are determined (rather
arbitrarily) by imposing the regularity condition
\be
\varphi_-^{(n)}(0)=\varphi_+^{(n)}(0),\qquad n=0,1,2 .
\ee
The result is 
\ba
c_2 &=& 0.225361 ,\\
c_3 &=& 0.398672 ,\\
c_4 &=& -2.77464 ,
\ea
which gives an approximate solution as regular as possible and asymptotically
exact. 

This procedure may be extended to the more general case of an arbitrary speed.
Since our interest is in the methodological viewpoint and in giving
specific examples we keep the above parameters 
($\alpha=2,\beta=\lambda=\epsilon=1$) but do not fix the speed. 
With these values, the travelling wave solutions of \eq{master}
obey
\ba
 2y^{iv} &-& 3vy^{'''}+(v^2-1) y^{''}+v y'= \\
        &=& -1+(y^{''}-v y'-1)^2(2y^{''}-v
 y'-y-2), 
\ea
and the fields are expressed by
\ba
 u &=& 1+v y'-y^{''} ,\\
 w &=& -2-y-vy'+2y^{''} .
\ea
The two fixed points correspond to the solutions
\be
y=-3,\qquad y=-\frac{\xi}{v} .
\ee
Perturbing around the first by setting $y=-3+\delta$ and expanding at the
first order in $\delta$ we obtain a linear differential 
equation with constant coefficients and characteristic polynomial
\be
\label{zeroes}
2p^4-3p^3v+(v^2-1)p^2+1=0 .
\ee
In \refapp{app:zeroes} we show that for each value of the speed $v$ 
the above equation has always two zeroes on the left complex half-plane and
the other two on the right.
Perturbing the second solution by setting $y=-\xi/v+\delta$ we find 
\be
p(p-v)(2p^2-pv-1)=0,\quad \rightarrow p=0,v,\frac{v\pm\sqrt{v^2+8}}{4} ,
\ee
and, for the same reasons as before, we are forced to match this boundary 
condition on the left at $\xi\to -\infty$ . 
Just to give a numerical result, the above procedure in the case $v=2$
provides the following approximate travelling wave
\be
y(\xi) = \left\{\begin{array}{ll}
y_-(\xi) & \xi < 0 , \\
y_+(\xi) & \xi \ge 0 , 
\end{array}\right.
\ee 
with 
\ba
y_- &=& c_1 e^{2\xi}+c_2 e^{\xi (1+\sqrt{3})/2}-\frac{\xi}{2}, \\
y_+ &=& -3 + (c_3 \cos\nu\xi+c_4\sin\nu\xi) e^{-\mu\xi},
\ea
where regularity up to the third order requires
\ba
c_1 &=&  0.255476 ,\\
c_2 &=& -0.690277 ,\\
c_3 &=& -1.217923 ,\\
c_4 &=&  2.565199 .
\ea

Turning to the numerical simulations, we fixed all the constants at the values
$\alpha = 2.0$, $\beta = 1.0$, $\epsilon = \lambda = 1.0$.
Assuming the equations of motion to be stabilized by the diffusion terms,
we have discretized them by a first order Euler scheme.

Concerning boundary conditions, they are easily dealt with at least in the
case of travelling wave solutions.
For generic positive speed $v$, the exact $u$ wave has 
asymptotically constant values whereas the exact 
$w$ wave tends to a constant on the 
right and to a linear function to the left.
Using these informations we may impose the correct 
boundary conditions minimizing finite size effects.
Anyway, we also checked independence of the results from the 
boundary conditions, at least when time is enough small for the size effects
not to be relevant.
For instance, if one exploits periodic boundary conditions 
and starts with the approximate (infinite volume) solution, then
a perturbation is seen to arise at the boundary and the simulation must 
be stopped when it collides with the localized reaction front of the $u$ wave.

Now, let us examine the output of the numerical simulation.
In \reffig{fig:wave1} we display the evolution of the approximate wave with
$v=2.0$. As one can see, it rapidly settles down to a stable wave 
travelling with the desired speed. 
In \reffig{fig:wave2} we repeat the simulation starting from the approximate wave with
speed $v=1.0$. In this case the approximate wave evolves toward an apparently
stable travelling wave with internal oscillations.
In \reffig{fig:wave3}, where $v=0.5$ is subcritical the same situation occurs.

These results suggest the following two scenarios below the critical wave 
speed: 
(i) there are no travelling waves or they are unstable; 
(ii) a stable travelling wave exists, but our approximate solutions are too
crude and their evolution does not converge to it.
Actually, in literature, we find examples of systems of this kind which
exhibit the scenario (i)~\cite{Gray}.

\section{Conclusions}
\label{sec:conclusions}

The results achieved in this paper show that the prolongation technique
reveals a fruitful tool to investigate morphogenesis and autocatalysis
modeling equations.

We have dealt with a class of RD systems including reaction terms which are
both quadratic and cubic in the fields. This class comprises well-known RD
equations such as the Gierer-Meinhardt models. The prolongation approach
allowed us to discover the existence of an algebraic structure inherent in
this kind of models. It turns out to be the algebra associated with the
similitude group. This feature could be important within a programme to
settle up a systematic of RD models, which might be classified according to
their algebraic properties.

The comparison between special solutions drawn from theory and the
corresponding ones obtained via numerical simulations is satisfactory. This
represents an encouragement to extend the algebraic strategy to handle more
complicated RD nonlinear evolution equations, keeping in mind cases in
two-space and one-time dimensions.

A direct search for particular solutions is possible as shown 
in~\cite{Ndayirinde} where the socalled $\tanh$ method is
utilized for a special cubic model. However, with this technique, 
calculations become rapidly cumbersome and are not suitable for model
classification. Just to give an example, we considered the cubic model
\bt{cubicsolvable}
u_t &=& \alpha u_{xx} -u^2 w + z_1 u + z_2 w + z_3 ,\\
w_t &=& \beta  w_{xx} -u^2 w + h_1 u + h_2 w + h_3 ,
\et
and looked for a travelling wave with speed $v$. In the case
$\alpha=\beta=v^2 z_1^2/(2 z_3^2)$, $h_1=z_1+z_2$, $h_2=0$, 
$h_3=z_3/z_1(z_1+z_2)$, we were able to find the solution
\ba
u &=&
\sqrt{z_1+z_2}\tanh\left(\frac{\xi}{v}\frac{z_3}{z_1}\sqrt{z_1+z_2}\right), \\
w &=& \frac{z_3}{z_1}+u,
\ea
which, in the special case $z_2=0$, 
reproduces one of the solutions in~\cite{Ndayirinde}.

To conclude our comments, we notice that mathematical models are only a
rough simplification of a complex reality where the mechanisms of the
biological, chemical and physical processes involved are often
unknown. Notwithstanding, we think that an important role of these models
is to get insight into possible interactions between specific processes and
to suggest sometime new experiments~\cite{Murray}.

\acknowledgements

\appendix

\section{Solution of a second order differential equation}
\label{app:mittag}

The solution of \eq{mittag} can be achieved via a trick by 
Mittag-Leffler~\cite{MittagLeffler}. In doing so, if $\tilde Y(\xi)$ is a
solution of the equation 
\be
\tilde Y''=a\tilde Y'-\frac{6}{25}a^2 \tilde Y+c \tilde Y^2,
\ee
with $a, c$ arbitrary constants, then
\be
H(\xi) = \left(\tilde Y'-\frac{2}{5}a \tilde Y\right)^2-\frac{2}{3} c
\tilde Y^3 ,
\ee
fulfills the equation 
\be
H' = \frac{6}{5}a H .
\ee
Thus, we can find a constant $H_0$ such that 
\be
\label{mittag1}
\left(\tilde Y'-\frac{2}{5}a\tilde Y\right)^2-\frac{2}{3} c\tilde Y^3 = H_0
\exp\left(\frac{6}{5}a\xi\right) . 
\ee
By making in~(\ref{mittag1}) the change of variables
\be
\tilde Y = \exp\left(\frac{2}{5}a\xi\right)\varphi,\qquad
z=\frac{5}{a}\sqrt\frac{c}{6}\exp\left(\frac{a\xi}{6}\right) ,
\ee
we get
\be
{\varphi'}^2(z) = 4\varphi^3(z)-k, \qquad k\ \mbox{constant},
\ee
wich is solved by 
\be
\varphi(z) = {\cal P}(z-z_0, 0, k).
\ee

\section{Prolongation of the quadratic model}
\label{app:quadratic}

In theory, quadratic reaction functions are compatible with
linearizability. However, algebraic constraints have to be satisfied. 
As we shall see, the quadratic Gierer-Meinhardt model does not fall into the
linearizable class, but showing this is not trivial. 

\subsubsection{The non-degenerate case: $\alpha+\beta \neq 0$}

Repeating the same computation done in the case of the non-degenerate cubic
model we find the following structure for the prolongation problem
\ba
y_x &=& F = \frac{1}{\alpha} a_1 u + \frac{1}{\beta} a_2 w + a_3 ,\\
y_t &=& G = a_1 u_x + a_2 w_x -\acomm 1 3 u - \acomm 2 3 w + a_4 ,
\ea
The reaction equation
\ba
F_u R_1 &+& F_w R_2 + \comm F {-\acomm 1 3 u - \acomm 2 3 w + a_4} = 0 ,\\
R_1 &=& \epsilon (uw-u) ,\ \ R_2 = \lambda (1-uw) ,
\ea
leads to the following algebra
\ba
\acomm 1 2 &=& 0 ,\\
\atcomm 1 1 3 &=& 0 ,\\
\atcomm 2 2 3 &=& 0 ,\\
\acomm 1 4 &=& \epsilon a_1 + \alpha \atcomm 3 1 3 ,\\
\acomm 2 4 &=& \beta\atcomm 3 2 3 ,\\
\acomm 3 4 &=& -\frac{\lambda}{\beta} a_2 ,\\
\frac{1}{\alpha}\atcomm 1 2 3 &+& \frac{1}{\beta}\atcomm 2 1 3 =
\frac{\epsilon}{\alpha} a_1 -\frac{\lambda}{\beta} a_2 .
\ea
Let us scale the generators
\be
a_1/\alpha \to a_1,\qquad a_2/\beta \to a_2 ,
\ee
in order to cast the incomplete algebra into the simpler form
\ba
\acomm 1 2 &=& 0 \label{inc1} ,\\
\atcomm 1 1 3 &=& 0 \label{inc2} ,\\
\atcomm 2 2 3 &=& 0 \label{inc3} ,\\
\acomm 1 4 &=& \epsilon a_1 + \alpha \atcomm 3 1 3 \label{inc4} ,\\
\acomm 2 4 &=& \beta\atcomm 3 2 3 \label{inc5} ,\\
\acomm 3 4 &=& -\lambda a_2 \label{inc6} ,\\ 
\atcomm 1 2 3 &+& \atcomm 2 1 3 =
\epsilon a_1 -\lambda a_2 \label{inc7} .
\ea
Since $\acomm 1 2 = 0$ we can write 
\be
\label{perm}
\atcomm 1 2 k = \atcomm 2 1 k , \qquad \forall k ,
\ee
and \eq{inc7} may be rewritten as
\be
\atcomm 1 2 3 = \atcomm 2 1 3 = \frac{1}{2}(\epsilon a_1 - \lambda a_2) .
\ee
Moreover, from \eq{inc4} and \eq{inc5} we find
\ba
\atcomm 2 1 4 &=& \alpha \comm{a_2}{\atcomm 3 1 3} ,\\
\atcomm 1 2 4 &=& \beta \comm{a_1}{\atcomm 3 2 3} .
\ea
Then, using again \eq{perm} we get
\be
\alpha \comm{a_2}{\atcomm 3 1 3} = \beta \comm{a_1}{\atcomm 3 2 3} ,
\ee
from which 
\ba
\lefteqn{
\alpha \comm{a_3}{\atcomm 2 1 3} -\alpha\comm{\acomm 1 3}{\acomm 2 3}
=} && \\
&& = \beta \comm{a_3}{\atcomm 1 2 3} -\beta\comm{\acomm 2 3}{\acomm 1 3},
\ea
or 
\be
\comm{\acomm 1 3}{\acomm 2 3} =
\frac{1}{2}\frac{\beta-\alpha}{\beta+\alpha}(\epsilon\acomm 1 3 - \lambda
\acomm 2 3) ,
\ee
by virtue of the Jacobi identity.
Now if we substitute this result into the Jacobi identity among 
the three operators
\be
a_1, \quad \acomm 1 3 , \quad \acomm 2 3 ,
\ee
we obtain $\atcomm 1 2 3 = 0$ and consequently
\be
\epsilon\ a_1 = \lambda\ a_2 .
\ee
This linear dependence between $a_1$ and $a_2$ constrains the incomplete
algebra to close collapsing onto $sim(2)$. In particular, complete
linearizability is not achieved.

\subsubsection{The degenerate case: $\alpha+\beta = 0$}

In the degenerate case, after proper rescaling, the incomplete algebra
turns out to be 
\ba
\acomm 1 4 &=& \atcomm 3 1 3 + \epsilon a_1 -\lambda a_0 ,\\
\atcomm 1 1 3 &=& 0 = \atcomm 2 2 3 ,\\
\acomm 2 4 &=& \atcomm 3 3 2 ,\\
\acomm 0 4 &=& -2 \atcomm 3 1 2 + \epsilon a_0 -\epsilon a_1 -\lambda a_2 ,\\
\atcomm 1 1 2 &=& \lambda a_0 ,\\
\atcomm 2 1 2 &=& \epsilon a_0 ,\\
\acomm 3 4 &=& \lambda a_2 ,\\
\acomm 0 i &=& 0, \qquad i=1,2,3 .
\ea
It is easy to show that $a_0$ must vanish. In order to see this we write
\be
\atcomm 1 0 4 = -2 \comm{a_1}{\atcomm 3 1 2}-\lambda\acomm 1 2 = 2 
\comm{\acomm 1 2}{\acomm 1 3}-\lambda \acomm 1 2 ,
\ee
but the left hand side vanishes because
\be
\atcomm 1 0 4 = \atcomm 0 1 4 -\atcomm 4 1 0 = \comm{a_0}{\atcomm 3 1 3 +
\epsilon a_1-\lambda a_0} = 0 .
\ee
Hence
\be
\comm{\acomm 1 2}{\acomm 1 3} = \frac{\lambda}{2}\acomm 1 2 ,
\ee
and
\ba
\lambda a_0 &=& \atcomm 1 1 2 = \frac{2}{\lambda} 
\comm{a_1}{\comm{\acomm 1 2}{\acomm 1 3}} = \nonumber \\
&=& \frac{2}{\lambda}\left\{
\comm{\acomm 1 2}{\atcomm 1 1 3}-\comm{\acomm 1 3}{\atcomm 1 1 2}
\right\} = \nonumber \\
&=& -\frac{2}{\lambda}\ \comm{\acomm 1 3}{a_0\lambda}= 0 .
\ea
The algebra simplifies and becomes
\ba
\acomm 1 4 &=& \atcomm 3 1 3 + \epsilon a_1  ,\\
\atcomm 1 1 3 &=& 0 = \atcomm 2 2 3 ,\\
\acomm 2 4 &=& \atcomm 3 3 2 ,\\
\atcomm 1 1 2 &=& 0 ,\\
\atcomm 2 1 2 &=& 0 ,\\
\acomm 3 4 &=& \lambda a_2 ,\\
\atcomm 3 1 2 &=& -\frac{1}{2}(\epsilon a_1 + \lambda a_2) \label{tmp10} .
\ea
Let us show that if 
the dimension of the space over which the above operator are defined
is finite, then 
\be
\label{tmp11}
\atcomm 1 1 2 = 0 = \atcomm 2 1 2 \Rightarrow \acomm 1 2 = 0,
\ee
from which, by using \eq{tmp10}, we obtain the desired constraint
\be
\epsilon a_1 + \lambda a_2 = 0 .
\ee 
The proof of \eq{tmp11} goes as follows: 
let $A,B,X$ be a triple of linear operators acting on a finite 
dimensional vector space, satisfying
\be
\comm A B = X, \ \ \comm A X = 0, \ \ \comm B X = 0 . 
\ee
Obviously $\mbox{Tr} X = 0$. Furthermore, since  
\be
\mbox{Tr} (A \comm B C) =  \mbox{Tr} (B\comm C A) ,   
\ee
we obtain, for $n>1$, 
\be
\mbox{Tr} X^n = \mbox{Tr} (X^{n-1}\comm A B) = \mbox{Tr}(A\comm B
{X^{n-1}}) = 0 .
\ee 
Thus, the characteristic polynomial of $X$ is 
\be
\det(X-\lambda) = (-\lambda)^{\dim X},
\ee
and $X$ has only the null eigenvalue. Let the canonical form of $X$ be
\be
\label{jordan}
\left(\begin{array}{ccc}
\Lambda_1 && \\
&\ddots&\\
&&\Lambda_N
\end{array}\right), 
\ee
(where $\Lambda_n$ are $n$-dimensional Jordan matrices).
The general form of the matrices which commute with~(\ref{jordan}) 
is block-rectangular where every block is a Toeplitz
upper triangular matrix~\cite{Gantmacher}. 
Then, it is easy to show that the condition 
$\comm A B = X$ implies that all the 
$\Lambda_n$ blocks are 1-dimensional and therefore $X=0$.

\section{The similitude group and Casimir operators}
\label{app:casimir}

Let us recall that given a Lie algebra $\cal L$ with generators $T_\alpha$
\be
\comm{T_\alpha}{T_\beta} = c^\gamma_{\alpha\beta} T_\gamma ,
\ee
a Casimir operator $\cal C$ is a polynomial in $\{T_\alpha\}$ which
commutes with all the generators
\be
\comm{{\cal C}}{T_\alpha} = 0 .
\ee
If the Lie algebra is semisimple, the metric tensor
\be
g_{\alpha\beta} = c^\tau_{\alpha\lambda} c^\lambda_{\beta\tau} ,
\ee
is non singular and the possible 
Casimir operators are contained in the sequence
\ba
I_2 &=& 
c^{\beta_2}_{\alpha_1\beta_1} 
c^{\beta_1}_{\alpha_2\beta_2} T^{\alpha_1} T^{\alpha_2} ,\\
I_3 &=& 
c^{\beta_2}_{\alpha_1\beta_1} 
c^{\beta_3}_{\alpha_2\beta_2}
c^{\beta_1}_{\alpha_3\beta_3}
 T^{\alpha_1} T^{\alpha_2} T^{\alpha_3} ,\\
&\cdots& 
\ea
where 
\ba
T^\alpha &=& g^{\alpha\beta} T_\beta ,\\
g^{\alpha\beta} g_{\beta\gamma} &=& \delta^\alpha_\gamma .
\ea
In the non semisimple case (for instance $sim(2)$), the independent 
Casimir operators must be built explicitly. Their number $N$ is 
expressed by 
\be
N = \dim {\cal L}-\max_{a_1,\cdots, a_{\dim {\cal L}}}\mbox{rank}\Vert
c^\rho_{\sigma\tau} a_\rho\Vert ,
\ee
according to the Beltrametti-Blasi theorem~\cite{Beltrametti}. 
In the particular case of $sim(2)$ we find $N=0$ and no Casimir
operator exists. On the other hand, the Euclidean subalgebra 
generated by $A_1$, $A_3$ and $A_5$ admits a Casimir operator
\be
{\cal C} = A_1^2 + A_5^2 ,
\ee
and since 
\be
\comm{{\cal C}}{A_4} = 2{\cal C} ,
\ee
it is easy to show that given a solution $y$ to the linearized problem
\ba
y_x &=& F_1 A_1 y + F_3 A_3 y , \\
y_t &=& G_1 A_1 y + G_5 A_5 y + \nu A_4 y ,
\ea
then the new pseudopotential 
\be
\tilde y = e^{-2\nu t}{\cal C} y,
\ee
is another solution corresponding to the same $(u,w)$ appearing in the
functions $F$ and $G$.

\section{Heuristic determination of {$\alpha^*$}}
\label{app:heuristic}

Let us look for a solution to the equation
\be
\theta' = -\theta+\xi\theta^2+\xi\theta^3 ,
\ee
with
\be
\theta(0) = \alpha>0 ,
\ee
in the following form
\be
\theta(\xi) = \sum_{n=1}^\infty e^{-n\xi} P_n(\xi),\qquad
P_1(\xi)=\alpha,\qquad P_{n<1}(\xi)\equiv 0 ,
\ee
where $P_n$ are functions to be determined.
Choosing the integration constants in a sensible way, we obtain
\be
P_n(\xi) = e^{(n-1)\xi}\int_{+\infty}^\xi e^{-(n-1)t}t \left\{
\sum_{m=1}^{n-1}P_m(t) P_{n-m}(t) + \sum_{m=1,l=1}^{n-1} P_m P_l P_{n-m-l}
\right\} .
\ee
The functions $P_n$ are polynomials, the first two being 
$P_1 = \alpha$, $P_2 = -\alpha^2(\xi+1)$ .
The initial value $\theta(0)$ is an unknown function of $\alpha$
admitting the series expansion
\be
\theta(0)=\sum_{n>1} P_n(0) = \alpha-\alpha^2+\frac 3 4
\alpha^3-\frac{17}{54}\alpha^4-\frac{605}{3456}\alpha^5 + \cdots = \sum
b_n\alpha^n .
\ee
This series shows a poor convergence near $\alpha\sim 1$, but in terms
of the mapped variable
\be
z = \frac{\alpha}{\alpha+1} ,
\ee
we find
\be
\theta(0) = z-\frac 1 4 z^3-\frac 7 {108} z^4 + \cdots = \sum b'_n z^n .
\ee
If we assume this series to have an infinite convergence radius and plot
the function 
\be
\sum b'_n \left(\frac{\alpha}{\alpha+1}\right)^n ,
\ee
we see that $\theta(0)$ appears to be upper bounded by a finite constant 
$\alpha^*$ whose numerical value can be estimated by truncating the
infinite coefficients $b'_n$. Taking the first 20 coefficients, we find 
\be
\label{alphastar}
\alpha^* = 0.756561 .
\ee

\section{Analysis of the roots of a quartic equation}
\label{app:zeroes}

If $Q(x)$ is a polynomial of degree $N$, then the number $Z$ of roots on 
the left side of the imaginary axis is given by
\be
Z = \frac{N}{2}+\frac{1}{2\pi}\int_{-\infty}^\infty\mbox{Re}\left. 
\frac{Q'(z)}{Q(z)}\right|_{z=iy} dy .
\ee
In the case $Q(x) = 2x^4-3vx^3+(v^2-1) x^2+1$ we have
\be
Z = 2 + \frac{1}{2\pi}\int_{-\infty}^\infty H(y) dy ,
\ee
where 
\be
H(y) = \frac{d}{dy}\left\{
\mbox{Arctan}\frac{3vy^3}{2y^4+(1-v^2)y^2+1}
\right\} .
\ee
Since $\mbox{Arctan}(\cdots)\to 0$ as $|y|\to 0$ we can conclude that
$Z=2$. 
The $\mbox{Arctan}$ argument is odd and its possible 
singularities give contributions which cancel in pairs.


\figlab{fig:4exact}{Exact solutions for the $u$, $w$ fields in the 
four cases corresponding to \eqs{CaseI}, (\ref{CaseII}), 
(\ref{CaseIII}) and (\ref{CaseIV}).
The numerical values of the parameters are respectively 
i)   $v=1$ , $\alpha_1=1$ , $\alpha_2=0$ , $\lambda_1=1$ , 
$\lambda_2=-1$ , $\mu_1=1$ , $\zeta=1$;
ii)  $v=1$ , $\alpha_1=1$ , $\alpha_2=-1$ , $\lambda_1=1$ , 
$\lambda_2=1$ , $\mu_1=1$ , $\zeta=1$;
iii) $v=1$ , $\alpha_1=2$ , $\alpha_2=1$ , $\lambda_1=1$ , 
$\lambda_2=-2$ , $\mu_1=1$ , $\zeta=0$;
iv)  $v=1$ , $\alpha_1=1$ , $\alpha_2=-1$ , $\lambda_1=1$ , 
$\lambda_2=1$ , $\mu_1=1$ , $\zeta=1$, $\rho_1=1$.
}
\figlab{fig:qualitative}{Qualitative behaviour of \eq{thetaevolution}.}
\figlab{fig:modified}{Analytical integrals of \eq{modified}.}
\figlab{fig:lambert}{Solution corresponding to Eqs~(\ref{lambert1})
and (\ref{lambert2}).}
\figlab{fig:homo}{Homogeneous solutions obtained from numerical integration
of \eqs{num1}.
We remark that from the value of $\alpha^*$ given in \eq{alphastar} we get 
$\alpha^*/(1+\alpha^*) \simeq 0.431$.}
\figlab{fig:wave1}{Evolution of the approximate travelling wave with $v=2.0$.
The values of the other parameters are $\alpha=2$, $\beta=1$, $\epsilon=1$, $\lambda=1$, $\Delta x=0.05$, $\Delta t=0.00025$, $\Delta x$ and $\Delta t$ being 
the space and time discretization steps.
}
\figlab{fig:wave2}{Like \reffig{fig:wave1}, but $v=1.0$.}
\figlab{fig:wave3}{Like \reffig{fig:wave1}, but $v=0.5$.}

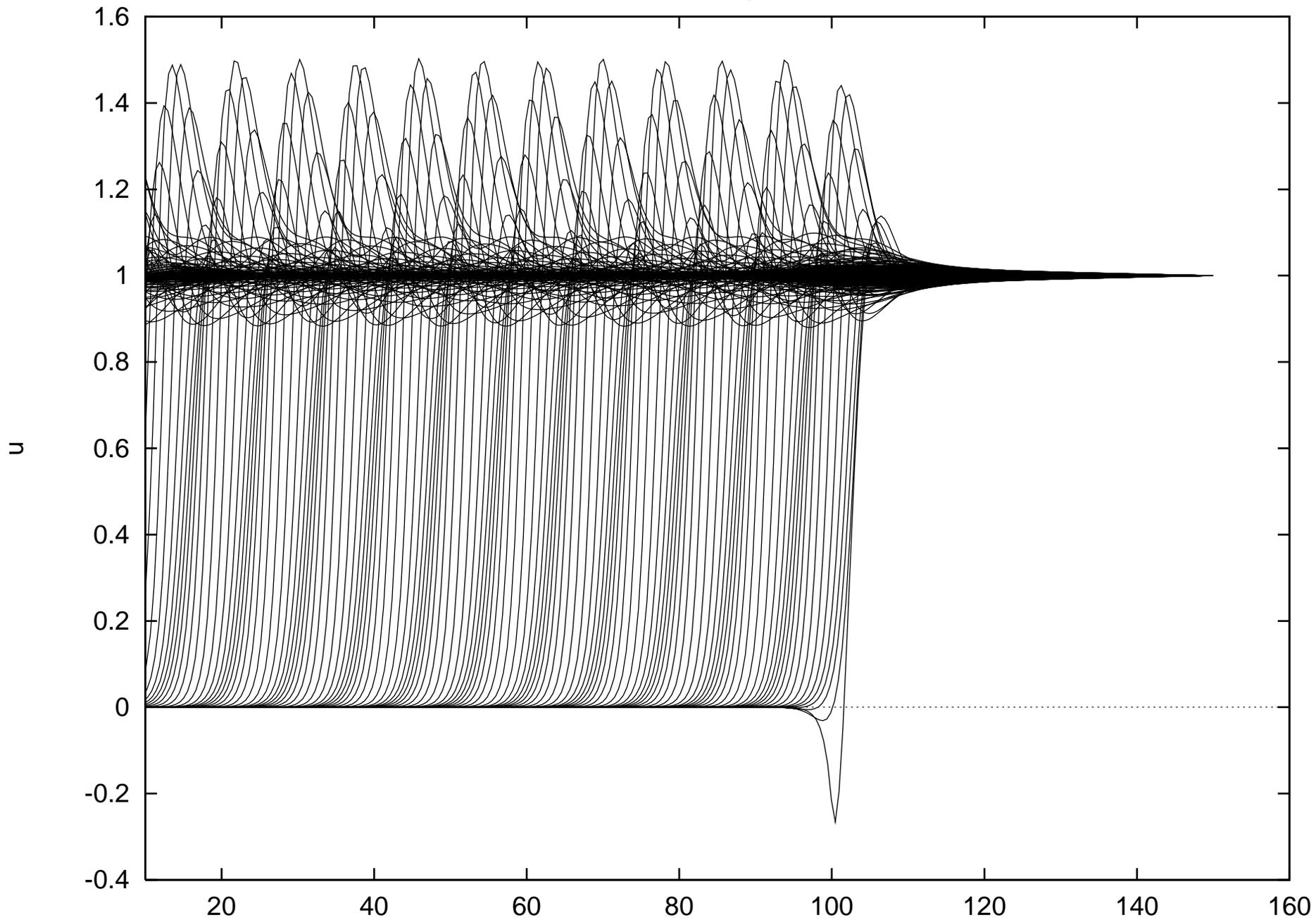
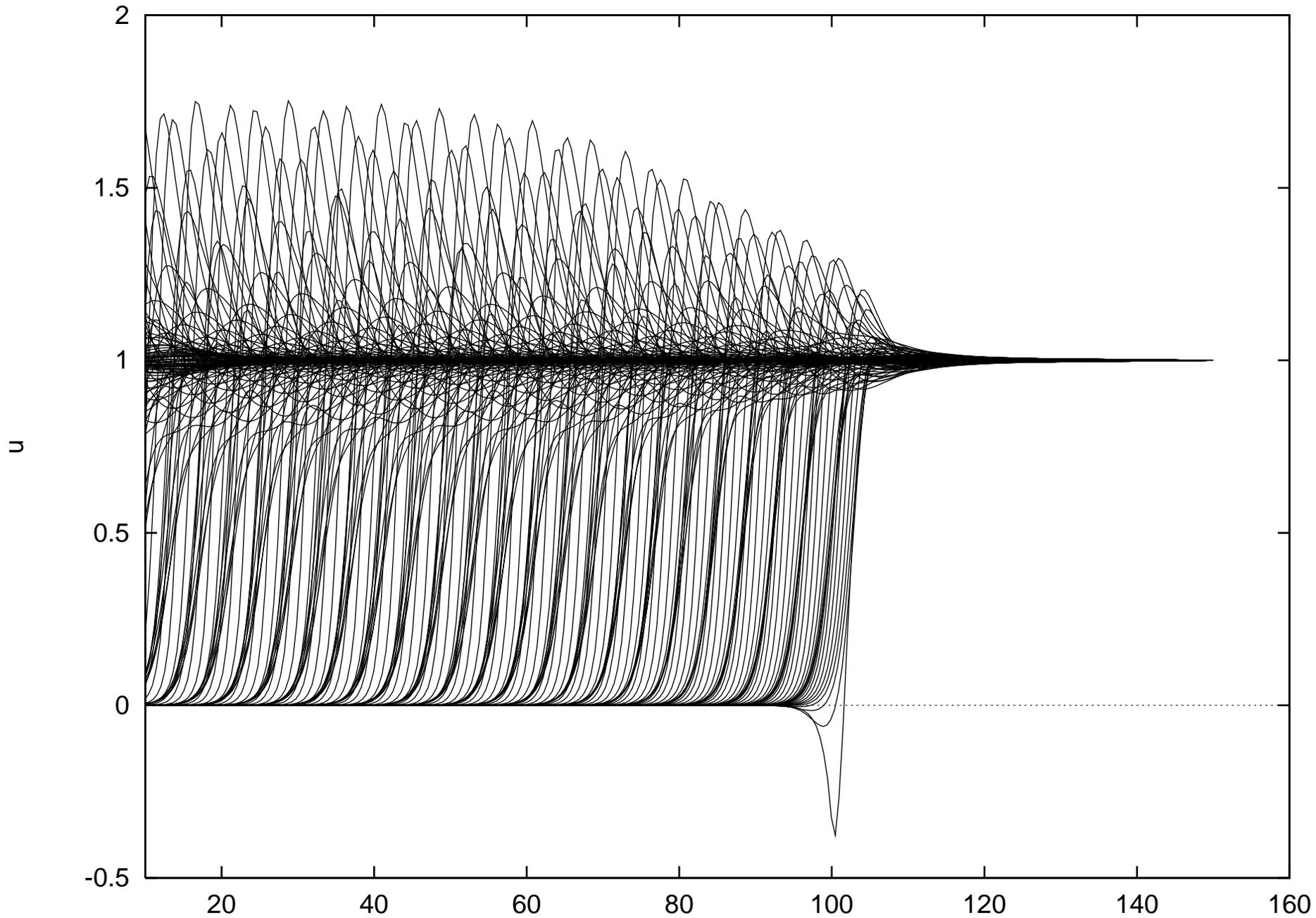
\end{document}